\documentclass[twocolumn,english,aps,prb,showpacs]{revtex4}
\usepackage[T1]{fontenc}
\usepackage[latin1]{inputenc}
\usepackage{float}
\usepackage{amsmath}
\usepackage{graphicx}
\usepackage{amssymb}

\makeatletter

\newcommand{\noun}[1]{\textsc{#1}}

\usepackage{graphicx}

\usepackage{babel}
\makeatother
\begin{document}

\title{Nodal liquid and $s$-wave superconductivity in transition
  metal dichalcogenides}

\author{B. Uchoa$^{1}$,  G. G. Cabrera$^{1}$, 
and A. H. Castro Neto$^{2}$}

\affiliation{$^{1}$ Instituto de F\'{\i}sica {}``Gleb Wataghin'' Universidade
Estadual de Campinas (UNICAMP), C. P. 6165, Campinas, SP 13083-970, Brazil \\
 $^{2}$ Department of Physics, Boston University, 590 Commonwealth
Ave., Boston, MA 02215}

\date{September 17, 2004}

\begin{abstract}
We explore the physical properties of a unified microscopic theory for
the coexistence of 
superconductivity and charge density waves in two-dimensional
transition metal dichalcogenides. In the case of particle-hole
symmetry the elementary particles are Dirac fermions at the nodes of
the charge density wave gap. When particle-hole symmetry is broken 
electron (hole) pockets are formed around
the Fermi surface. The superconducting ground state emerges
from the pairing of nodal quasi-particles mediated by acoustic phonons
via a piezoelectric coupling. We calculate several properties in the
$s$-wave superconducting phase, including specific heat, ultra-sound
absorption, nuclear magnetic relaxation, thermal, and optical
conductivities. In the case with particle-hole symmetry, the specific heat
jump at the transition deviates strongly from ordinary superconductors.
The NMR response shows an anomalous anisotropy due to the broken time-reversal 
symmetry of the superconducting gap, induced by the triple CDW state.
 The loss of lattice inversion symmetry in the charge density wave phase leads
to anomalous coherence factors in the optical conductivity and to
the appearance of an absorption edge at the optical gap energy. Furthermore,
optical and thermal conductivities display anomalous peaks in the
infrared when particle-hole symmetry is broken.
\end{abstract}


\maketitle

\section{Introduction}

The quasi two-dimensional (2D) transition metal dichalcogenides (TMD)
2H-TaSe$_{2}$, 2H-TaS$_{2}$ and 2H-NbSe$_{2}$ are layered compounds
where $s$-wave superconductivity coexists with a charge density
wave (CDW) \cite{Withers,Wilson} at low temperatures, 
and whose transport properties are highly anisotropic in the 
high temperature CDW phase \cite{transport}. There is
a vast literature reporting anomalous effects in the CDW phase,
including, non-linear Hall effect, anomalous impurity effects in the
superconducting (SC) phase \cite{hall}, stripe phases \cite{stripes},
and different regimes of commensurability {\small \cite{McWhan}}. 
Recent angle resolved photoemission experiments (ARPES) 
reveal that the quasi-particles of TaSe$_{2}$ have a marginal
Fermi liquid (MFL) lifetime \cite{valla3}. This scenario becomes more
exciting by the verification that some of the physical properties
of TMD, such as the linear growth of the normal resistivity with 
temperature \cite{transport}, 
and the strong anisotropy in the in-plane and out-of-plane transport
are similar to the same properties in the high temperature superconductors (HTc).
HTc do not show a CDW gap but a $d-$wave \emph{pseudo-gap} 
coexisting with the superconducting phase. In both
cases, the transport and thermodynamic properties are weakly dependent
on the application of external fields in the normal/pseudo-gap phase,
and strongly dependent on them in the superconducting phase \cite{Klemm}. 
Furthermore,
the application of pressure in TMD favors the superconductivity 
and suppresses the CDW phase \cite{chu}, in a close
analogy with the HTc phase diagram of temperature versus doping. 
Differently from the HTc, however, the TMD are very clean materials. 
The anomalous TMD properties are sample independent and can help to clarify the
physics behind a whole class of low-dimensional superconductors. 

The interpretation of the experimental data in 
TMD is however still very controversial.
Within the Peierls theory, the CDW gap formation in 1D systems is
usually due to nested Fermi surfaces. In 2D systems, the nesting is
not perfect and some parts of the Fermi surface may not be gaped.
Early band structure calculations \cite{Wexler} indicated that
the $\Gamma$ centered sheets ($S_{I}$) are nested with the $K$
centered ones (S$_{II}$) by the $\mathbf{Q}_{i}$ ($i=1,2,3$) wavevectors
of the triple-CDW (see Fig. 1). The value of the CDW wave-vector, 
$|\mathbf{Q}_{i}|\sim\frac{1}{3}\Gamma K$, 
measured by neutron diffraction \cite{neutrons,Wilson2}, and recent
scanning tunneling microscopy (STM) experiments \cite{sacks, straub, tonjes}  
confirm the plausibility of a nesting scenario. An alternative theory
proposed by Rice and Scott \cite{rice} is based on a Fermi surface independent
CDW mechanism, where the CDW wavevectors connect the saddle points (indicated
in Fig. 1 around $\frac{1}{2}\Gamma K$) of the transition metal
$d-$bands, generating a logarithmic divergence in the electronic
susceptibility. However, the
saddle point energy in NbSe$_{2}$ is too large ($\sim50$ meV) in comparison to the CDW
ordering thermal energy $k_{B}T_{CDW}\sim3$ meV to allow a saddle point driven
instability \cite{Rossnagel}. In TaSe$_{2}$, however, ARPES 
has observed an extended saddle band along $\Gamma K$. This band
is nearly flat and closer to the Fermi energy than the band calculations
predicted \cite{seifarth, liu}. As the saddle points are not well defined
in this case, it is questionable to justify the CDW wave-vector measured
with neutrons by some mechanism related to special parts of the saddle
bands. More experimental studies are required to elucidate this point.

If on one hand these arguments seem to rule out at least a conventional
saddle point mechanism, consensus on the origin of the CDW instability
has not been reached. STM scans at 4.2 K in TaSe$_{2}$, TaS$_{2}$
and NbSe$_{2}$ show that the amplitude of the CDW gap is 
$\Delta_{CDW}\sim80,\,50$, and 34 meV, respectively \cite{chen}. 
The ability of ARPES to measure
the superconducting gap $\Delta_{s}\sim1$ meV \noun{$\ll\Delta_{CDW}$} 
in NbSe$_{2}$\noun{,}
combined with the complete failure of ARPES
to detect traces of the CDW gap in the Brillouin zone of TaSe$_{2}$
and NbSe$_{2}$ \cite{Valla2,valla3} were interpreted as an evidence
that the Fermi surface is weakly covered by the CDW. We observe that
the ARPES results seem to be in contradiction with the STM
data, and cannot explain the non-Fermi liquid transport in the TaSe$_2$
crystal. One possibility is that the ARPES data are obscured by the
strong dependence of CDW gap with the directions of the Brillouin
zone combined with the formation of pockets in the points of the Fermi
surface where $\Delta_{c}(\mathbf{k})=0$ 
($\textrm{max}${[}$\Delta_{c}(\mathbf{k})${]} $=\Delta_{CDW}$).
Another possibility is that the ARPES electronic dipole matrix elements 
vanish for certain states in the CDW phase due to the broken spacial 
inversion symmetry (detected in neutron scattering \cite{neutrons}) 
forbidding the observation of some bands.

The strong resemblance of the normal CDW phase resistivity of TaSe$_{2}$
with the HTc \cite{vescoli} and the anomalous quasi-particle
life-time, given by the inverse of the imaginary part of the
electronic self-energy \cite{valla3}
$\textrm{Im}\,\Sigma(k_{F},\omega)\propto\tau_{0}^{-1}+b|\omega|$, 
indicates that a marginal Fermi liquid theory
\cite{littlewood} should be developed as the basis of a minimal model
unifying the CDW and superconducting phases in TaSe$_{2}$. 
The experimental verification
that $k_{B}T_{CDW}\ll\Delta_{CDW}$ for all the TMD (in TaSe$_{2}$
for example, $k_{B}T_{CDW}\sim120\,\textrm{K} \sim 12$ meV) gives a good
indication that a strong coupling CDW theory is required. 

One of us (A.H.C.N.) \cite{Neto} has recently proposed a unified picture 
for the CDW and SC phases where
the elementary particles are Dirac fermions that are created
in the region where the CDW gap vanishes, leading to the generation of 
a nodal liquid. According to neutron diffraction studies, the inversion center
of the crystal is lost in the CDW phase \cite{neutrons}, allowing for
the possibility of piezoelectric effects. In a system with nodal 
quasiparticles, the piezoelectric coupling is a marginal coupling from
the renormalization group (RG) point of view, while the usual electron-phonon 
coupling is irrelevant under the RG \cite{subir}.
Based on a \emph{tight-binding} description of the electronic orbitals 
\cite{Varma}, and on the
assumption of imperfect nesting between different Fermi surface sheets,
the model of ref.~[\onlinecite{Neto}] proposes a \emph{}$f$-wave symmetry CDW gap, 
with lobes
along the saddle point directions, and six nodes at the points where
the gap is zero (see Fig. 1). The proposed  CDW gap is odd in the Brillouin
zone, due to the symmetry of the electron-phonon coupling \cite{Neto}, and 
due to the absence of inversion symmetry in the CDW phase, changing sign
in each node. The superconductivity emerges from Cooper pairing between
the Dirac fermions mediated by acoustic phonons via a piezoelectric
coupling. We propose that the Fermi surface is fully gaped by the
superposition of the CDW and the $s-$wave superconducting (SC) order 
parameters. This
model is able to correctly explain some of the anomalous properties
of the TMD like the marginal quasi-particles life-time in TaSe$_{2}$,
the dependence of the normal-superconducting phase transition with the lattice
parameters, and the metallic behavior of the resistivity in the CDW
phase \cite{Neto}. 

The geometry of the proposed CDW gap is similar to the Brillouin zone
of graphite, where the nodes represent the points where the conduction
and valence $\pi-$bands cross each other \cite{mele}. In contrast 
to graphite, the lattice inversion symmetry is broken in the distorted
phase, and the piezoelectricity can arise. As it is usually observed
in insulators, since metals screen the polarization fields, one
may ask: is actually possible to find piezoelectricity in a superconductor?
To answer this question, we should consider first that in a nodal liquid 
the density of states (DOS) goes to zero in the nodes, and therefore the
electrons cannot effectively screen electric fields. Hence, one can
conciliate a metallic theory (with gapless quasi-particle excitations)
with piezoelectricity. The rigorous vanishing of the DOS at the Fermi surface,
however, is not essential for the piezoelectricity to appear.  It
is sufficient to consider that the electrons of low lying momentum 
(for example, in a small pocket around the nodes) are "slow" enough to couple with the 
acoustic phonons of the polarized lattice. 

\begin{figure}
\includegraphics[%
  scale=0.43]{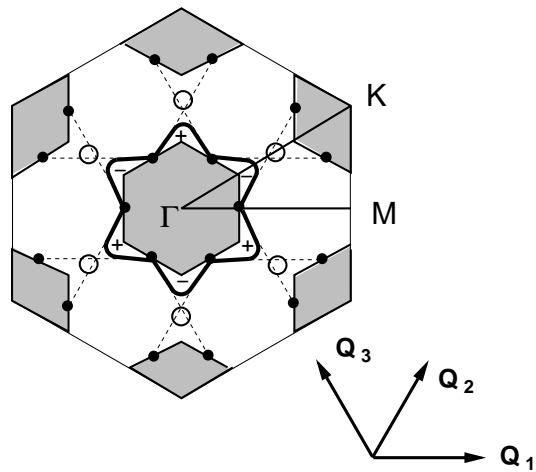}

\caption{{\small Schematic representation of the TMD Fermi surface. The $\Gamma$
centered sheet ($S_{I}$) is nested with the $K$ centered ones ($S_{II}$)
by the CDW wavevectors $\mathbf{Q}_{i}$. A CDW gap develops in the
two sheets, except in the nodal points, indicated by the black filled
circles. The empty circles are the saddle points. The thick solid
line around the $S_{I}$ sheet is the proposed CDW gap. The dashed
lines indicate the nodes connected by $\mathbf{Q}_{i}$.}}
\end{figure}

If the piezoelectricity and the metallic character are not mutually
excluding, it remains the question of how the polarization vector
affects the phase coherence of the condensate. The answer to this
question can be found in the collective modes. The electromagnetic
gauge invariance of the superconducting state 
is provided by the longitudinal
response of the collective excitations, that screen the electrons
through a cloud of virtual plasmons \cite{Pines}. Only
the plasmons respond to the longitudinal fields and give rise to
screening. Since the piezoelectricity involves electric fields only,
it does not affect the phase coherence of the electrons. In a previous
work \cite{Uchoa}, we have shown by means of a semi-classical calculation
that piezoelectricity is not only consistent with the stability
of the condensate as it is possibly behind the quantum critical points
(QCP) observed experimentally in the TaSe$_{2}$ phase diagram, separating
the $T=0$ commensurate phases from the stripe phase as a function of
the applied pressure.

The organization of the paper is as follows: in sec. II we introduce the
model Hamiltonian of the CDW and superconducting phase; 
in sec. III we derive
the superconducting gap equation; sec. IV is devoted to the thermodynamics 
of the superconducting phase while in sec. V we calculate the acoustic 
attenuation rate and the NMR response; in sec. VI we calculate the optical 
and thermal conductivities; in sec. VII we discuss the Meissner effect; sec. VIII
contains our conclusions.

\section{The Hamiltonian}

The nodal system is composed of two subsystems defined by the nodes
of the CDW state 
which are connected by the triple-CDW wavevectors $\mathbf{Q}_{i}$
($i=1,2,3$). It is convenient to introduce the spinors 
\begin{eqnarray*}
\Psi_{i,\sigma}(\mathbf{k}) & = & \left(
\begin{array}{c}
c_{\mathbf{k},\sigma}
\\
c_{\mathbf{k}+\mathbf{Q}_{i},\sigma}
\end{array}\right)=\left(
\begin{array}{c}
\psi_{+,i,\sigma}(\mathbf{k})
\\
\psi_{-,i,\sigma}(\mathbf{k})
\end{array}\right)\,,
\end{eqnarray*}
where $+,-$ indicate the two nodal spaces, where
$c_{\mathbf{k},\sigma}^{\dagger}(c_{\mathbf{k},\sigma})$ are creation
(annihilation) operators for electrons with momentum \textbf{$\mathbf{k}$}
and spin $\sigma=\uparrow,\downarrow$. The electronic Hamiltonian
in the normal CDW phase is composed of two terms, \[
H_{CDW}=H_{e}+H_{e-c}\,.\]
$H_{e}$ is the Hamiltonian of the free electrons in
the vicinity of the nodes, 
\begin{eqnarray}
H_{e} & = & \sum_{\mathbf{k},\sigma,i}\left[\epsilon_{\mathbf{k}}\,
  c_{\sigma,\mathbf{k}}^{\dagger}c_{\sigma,\mathbf{k}}+\epsilon_{\mathbf{k}+\mathbf{Q}_{i}}\, c_{\sigma,\mathbf{k}+\mathbf{Q}_{i}}^{\dagger}c_{\sigma,\mathbf{k}+\mathbf{Q}_{i}}\right]\nonumber 
\\
 & = &
  \frac{1}{2}\sum_{\mathbf{k},a,b,\sigma,i}\:\psi_{a,i,\sigma}^{\dagger}(\mathbf{k})\left[(\epsilon_{\mathbf{k}}+\epsilon_{\mathbf{k}+\mathbf{Q}_{i}})\eta_{0}^{a\,
  b}\right.\nonumber 
\\
 &  &
 \qquad\qquad\quad\left.+(\epsilon_{\mathbf{k}}-\epsilon_{\mathbf{k}+\mathbf{Q}_{i}})\eta_{3}^{a\,
 b}\right]\psi_{b,i,\sigma}(\mathbf{k}),
\label{He}
\end{eqnarray}
where $\eta_{\nu}$ ($\nu=0,1,2,3$) are Pauli matrices that act
in the nodal indexes $a,\, b=\pm$, and $\epsilon_{\mathbf{k}}$ is
the free electron dispersion.
In our convention, $\eta_{0}$ is the identity and $\nu=1,2,3$
indexes the $x,\, y,\, z$ directions, respectively. The second term
in the Hamiltonian, $H_{e-c}$, is the CDW exchange Hamiltonian between 
electrons  situated
in two different nodes connected by $\mathbf{Q}_{i}$, 
\begin{eqnarray}
H_{e-c} & = & \sum_{i,\mathbf{k}}\,\Delta_{c\,\mathbf{k}}\, c_{\mathbf{k}\sigma}^{\dagger}c_{\mathbf{k}+\mathbf{Q}_{i}}+h.c.\nonumber \\
 & = &
 \sum_{i,\mathbf{k},\sigma,a,b}\:\Delta_{c\,\mathbf{k}}\,\psi_{a,i,\sigma}^{\dagger}(\mathbf{k})\eta_{1}^{a\,
 b}\psi_{b,i,\sigma}(\mathbf{k})\,.
\label{He-c}
\end{eqnarray}
where $\Delta_{c\mathbf{k}}$ is the CDW gap, with odd-parity
in the nodal space due to the loss of the lattice inversion symmetry.
This term arises from the scattering of the electronic
wave function by the CDW periodic superstructure. 

Applying the \emph{nesting} condition $\epsilon_{\mathbf{k}}+\epsilon_{\mathbf{k}+\mathbf{Q}}=0$,
(see Fig. 2) in Eq. (\ref{He}), and taking the long-wavelength, low-energy limit,
the Hamiltonian in the CDW phase reads, 
\begin{equation}
H_{CDW}=\sum_{\mathbf{k},\sigma,i}\:\Psi_{i,\sigma}^{\dagger}(\mathbf{k})\left[v_{F}k_{\perp}\eta_{3}+v_{\Delta}k_{\parallel}\eta_{1}\right]\Psi_{i,\sigma}(\mathbf{k}),
\quad
\label{HCDW}
\end{equation}
where $k_{\perp}$ and $k_{\parallel}$ are
the momentum components in the normal and parallel directions to the
Fermi surface, respectively, $v_{F}$ is the Fermi surface velocity
and, $v_{\Delta}=\frac{\partial\Delta_{c}}{\partial k_{\parallel}}$.
The CDW elementary excitations around the nodes are therefore fermions
which follow the two-dimensional massless Dirac Hamiltonian, 
similarly to the two-band electronic description of graphite \cite{mele}. 

\begin{figure}
\begin{center}\includegraphics[%
  scale=0.38]{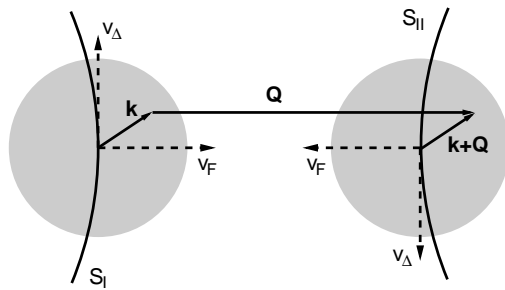}\end{center}

\caption{{\small Nesting condition in the two sheets $S_{I}$ and $S_{II}$
of the TMD Fermi surface. The} \textbf{\small }{\small momentum} \textbf{\small $\mathbf{k}$}
{\small outside $S_{I}$ is mapped by a CDW wavevector displacement
into $\mathbf{k}+\mathbf{Q}$, inside S$_{II}$. As the free electron
dispersion $\epsilon_{\mathbf{k}}$ is odd with respect to the Fermi
surface sheets, we have $\epsilon_{\mathbf{k}}=-\epsilon_{\mathbf{k}+\mathbf{Q}}$. }}
\end{figure}

The broken lattice inversion symmetry due to the CDW gap allows the
piezoelectricity in the crystal. We propose that the electron-phonon
coupling is piezoelectric, giving rise to a pairing of Dirac
fermions in the singlet state through the triple-CDW superstructure.
In contrast with usual Cooper pairs, whose electrons are paired
across the Fermi surface, these pairs are formed by electrons located
in different nodes linked by a CDW wavevector \textbf{$\mathbf{Q}_{i}$}
(see Fig. 1). The pairing approximation consists in assuming
a condensate of pairs whose center of mass have momentum $\mathbf{Q}_{i}$
and zero spin. This assumption clearly \emph{violates} the time-reversal
symmetry of the superconductor order parameter $\Delta_{s}$. According
to Anderson \cite{Anderson}, the strong insensitivity of the BCS
superconductors to impurities 
is due to the tendency of electrons to be in the state of highest
possible degeneracy in the condensate, implying pairing each electron
with its symmetric in spin and momentum. In such case, the scattering
channels promote transitions between two degenerated states, keeping the 
system coherent. The absence of time reversal symmetry should destroy
the condensate in the presence of a very small impurity concentration
\cite{Anderson84}. In the case TMD, however, the CDW scattering does
not affect the degeneracy of the condensate as far as the Dirac fermions
$\Psi_{i}$ living in different nodal subspaces (indexed by the three
CDW directions $i=1,2,3$) remain decoupled. For this reason, we may
drop the $i$ index from now on. 

After tracing the phonons, the piezoelectric pair-interaction has
the form \cite{Neto}, 
\begin{eqnarray*}
H_{P} & = &
-g\!\sum_{\mathbf{k},\mathbf{k}^{\prime}}\sum_{a,b,c,d}\,\eta_{2}^{a\,
  b}\eta_{2}^{c\,
  d}\,\psi_{a,\uparrow}^{\dagger}(\mathbf{k})\,\psi_{b,\downarrow}^{\dagger}(-\mathbf{k})
\\
 &  &
 \qquad\qquad\qquad\times\psi_{c,\uparrow}(\mathbf{k}^{\prime})\,\psi_{d,\downarrow}(-\mathbf{k^{\prime}})\,,
\end{eqnarray*}
where $g$ is the coupling
constant. The choice of the anti-symmetric
Pauli matrix $\eta_{2}$ incorporates the broken symmetry of the
superconducting gap. In the mean field approximation, the pairing Hamiltonian
reads,
\begin{eqnarray}
H_{P} & = &
\sum_{\mathbf{k}}\sum_{a,b}\left[\Delta_{s}\psi_{a,\uparrow}^{\dagger}(\mathbf{k})\,\eta_{2}^{a\,
    b}\psi_{b,\downarrow}^{\dagger}(-\mathbf{k})+h.c.\right]+\,\,\frac{\Delta_{s}^{2}}{g}\nonumber 
\\
\label{pairing}
\end{eqnarray}
where 
\begin{equation}
\Delta_{s}=-g\sum_{\mathbf{k}}\sum_{a,b}\,\langle\psi_{a,\uparrow}(\mathbf{k})\,\eta_{2}^{a\,
  b}\psi_{b,\downarrow}(-\mathbf{k})\rangle
\label{Delta}
\end{equation}
is the complex superconductor order parameter. 

So far, we have discussed the problem with particle-hole symmetry, that is,
the chemical potential $\mu$ is exactly at the Dirac point ($\mu=0$).
In order to include the situation where particle-hole symmetry is broken we
have add 
to Eq. (\ref{HCDW}) a chemical potential term 
\begin{equation}
H_{\mu} = -\mu\,\sum_{\sigma,a}\psi_{a,\sigma}^{\dagger}(\mathbf{k})\psi_{a,\sigma}(\mathbf{k})\,.
\label{Hmu}
\end{equation}
This term introduces an electron ($\mu >0$) or hole ($\mu<0$) pocket around
the Dirac point producing a finite density of states. 

In order to diagonalize the problem it is convenient to extend the spinorial notation
to the Nambu space
\begin{equation}
\Psi(\mathbf{k})=\left(\begin{array}{c}
\psi_{+,\uparrow}(\mathbf{k})\\
\psi_{+,\downarrow}^{\dagger}(-\mathbf{k})\\
\psi_{-,\uparrow}(\mathbf{k})\\
\psi_{-,\downarrow}^{\dagger}(-\mathbf{k})
\end{array}\right)\,,
\label{spinor}
\end{equation}
with $\mathbf{k}$ defined with respect to the nodes. We
introduce a new set o Pauli matrices $\tau_{\mu}$ which operates
in the space $(\uparrow{k},\downarrow-{k})$. 
Denoting $\tau_{\mu}\eta_{\nu}$ as the tensor product between the
Nambu and nodal spaces, it is not difficult to see that the full Hamiltonian
is written as
\begin{eqnarray}
H & = & \sum_{\mathbf{k}}\,\Psi^{\dagger}(\mathbf{k})\left[v_{F}k_{\perp}\tau_{0}\eta_{3}+v_{\Delta}k_{\parallel}\tau_{0}\eta_{1}\right.\nonumber \\
 &  &
 \qquad\qquad\qquad\left.+\Delta_{s}\tau_{1}\eta_{2}-\mu\tau_{3}\eta_{0}\right]\Psi(\mathbf{k})\,.
\label{H1}
\end{eqnarray}
Notice that the gauge symmetry of the problem $\psi\to\psi e^{i\theta}$,
and $\Delta_{s}{e}^{2i\theta}\to\Delta_{s}$, is broken at the mean-field
level.  With this notation, the SC order parameter is given by :
\begin{eqnarray}
\Delta_{s} & = &
-g\sum_{\mathbf{k}}\,\langle\Psi^{\dagger}(\mathbf{k})\,\tau_{1}\eta_{2}\Psi(\mathbf{k})\rangle\,.
\label{gapPsi}
\end{eqnarray}
The diagonalization of the Hamiltonian leads to four branches
of excitations:
\begin{eqnarray}
\pm E_{\mathbf{k},\pm\mu} & \equiv &
\pm\sqrt{(v_{F}\bar{k}\pm\mu)^{2}+\Delta_{s}^{2}}\,,
\end{eqnarray}
where $\bar{\mathbf{k}}=\vec{k}_{\perp}+(v_{\Delta}/v_{F})\vec{k}_{\parallel}$
is the in-plane anisotropic momentum, with $\bar{k}\equiv|\bar{\mathbf{k}}|$.
In the normal phase, we identify two branches of excitations (we assume $\mu>0$):
\begin{eqnarray*}
\pm E_{\mathbf{k},\pm\mu} & \stackrel{\Delta_{s}\to0}{\longrightarrow} & \left\{ \begin{array}{ll}
\pm v_{F}\bar{k}+\mu\, & \quad\:\textrm{(hole-like branch)}\\
\\\pm v_{F}\bar{k}-\mu\, & \quad\textrm{ (particle-like
  branch)}\,,\end{array}\right.
\end{eqnarray*}
which are related to hole and particle-like pockets around the CDW nodes 
(for $\mu<0$, the nomenclature is exchanged). 
The two branches are physically equivalent to each other, except for a 
constant equal to $-\sum_{\mathbf{k}}2\mu$, 
integrated in the volume of the Dirac cone.
The optical gap in the SC phase is 
$2\sqrt{\mu^{2}+\Delta_{s}^{2}}$, as one can see from Fig. 3.

\begin{figure}[H]
\begin{flushleft}$\!\!\!\!\!\!\!$\includegraphics[%
  scale=0.9]{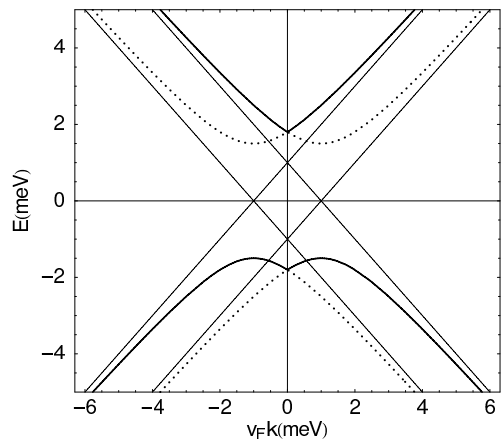}\vspace{-0.3cm}\end{flushleft}

\caption{{\small Dirac fermion dispersion  in the pocket with the opening
of the SC gap for $|\mu|=\frac{3}{2}\Delta_{s}=1$ meV. Each band has
two pocket branches indicated by the dotted and thick solid lines. The thin solid lines with
the vertex above (below) the Fermi energy $E=0$ represent the hole (particle)-like branches of the
Dirac cone in the normal CDW phase.}}
\end{figure}

\section{The gap equation}

To calculate the gap self-consistently, we use the standard many-body
Green's function method. Since Hamiltonian (\ref{H1}) has a quadratic
form,
$H=\sum_{\mathbf{k}}\Psi^{\dagger}\stackrel{\leftrightarrow}{\omega}\Psi$
, its corresponding Green function in the 4$\times4$ space is: 
\[
\stackrel{\leftrightarrow}{G}(\mathbf{k},i\omega_{n})=-\int_{0}^{\beta}\textrm{d}\tau\,\textrm{e}^{i\omega_{n}\tau}\langle T_{\tau}[\Psi\Psi^{\dagger}]\rangle=\left(i\omega_{n}-\stackrel{\leftrightarrow}{\omega}_{\mathbf{k}}\right)^{-1}\,,\]
 where $T_{\tau}$ is the time-ordering operator in imaginary time,
$\omega_{n}$ are the fermionic Matsubara frequencies, $\beta=1/(k_{B}T)$
is the inverse of temperature, $k_{B}$ is the Boltzmann constant,
and 
\begin{eqnarray}
\stackrel{\leftrightarrow}{\omega}_{\mathbf{k}} & \equiv &
v_{F}\tau_{0}\vec{\eta}\cdot\bar{\mathbf{k}}+\Delta_{s}\tau_{1}\eta_{2}-\mu\,\tau_{3}\eta_{0}\,,
\end{eqnarray}
is the dispersion tensor with $v_{F}\vec{\eta}\cdot\bar{\mathbf{k}}\equiv v_{F}k_{\perp}\eta_{3}+v_{\Delta}k_{\parallel}\eta_{1}$.
Exploring the anti-commutative property of the Pauli matrices, the
Green function which is systematically used in our calculation is:
\begin{equation}
\stackrel{\leftrightarrow}{G}(\omega_{n},\mathbf{k})=-\left(i\omega_{n}+\stackrel{\leftrightarrow}{\omega}_{\mathbf{k}}\right)\frac{\omega_{n}^{2}+E_{\mathbf{k}}^{\prime\,2}+2\mu
  v_{F}\tau_{3}\vec{\eta}\cdot\bar{\mathbf{k}}}{\left[\omega_{n}^{2}+E_{\mathbf{k},\mu}^{2}\right]\!\!\left[\omega_{n}^{2}+E_{\mathbf{k},-\mu}^{2}\right]} ,
\label{GreenF1}
\end{equation}
where \[
E_{\mathbf{k}}^{\prime\,2}\equiv v_{F}^{2}\bar{k}^{2}+\mu^{2}+\Delta_{s}^{2}\,=\, E_{\mathbf{k},\pm\mu}^{2}-2v_{F}(\pm\mu)\bar{k}\,.\]

Noting that $\langle\Psi_{\alpha}^{\dagger}(\mathbf{k})\Psi_{\beta}(\mathbf{k})\rangle$
is the retarded part of the Green function, $G_{\beta\,\alpha}(\mathbf{k},\tau\to0_{-})$,
we see from Eq. (\ref{gapPsi}) that the amplitude of the mean-field
gap is written in the Nambu notation as
\begin{eqnarray*}
2\Delta_{s} & = &
-\frac{{g}}{\beta}\sum_{\mathbf{k}}\sum_{\omega_{n}=-\infty}^{\infty}\,\textrm{Tr}\left[\tau_{1}\eta_{2}\stackrel{\leftrightarrow}{G}(\omega_{n},\mathbf{k})\right]\,.
\end{eqnarray*}
Evaluating the trace yields:\begin{eqnarray}
2\Delta_{s} & = & \frac{{g}v_{F}\Delta_{s}}{2\pi
  v_{\Delta}}\sum_{\sigma=\pm1}\int_{0}^{\Lambda}\textrm{d}\bar{k}\,\frac{\bar{k}}{E_{\mathbf{k},\sigma\mu}}\tanh\left(\beta\frac{E_{\mathbf{k},\sigma\mu}}{2}\right)\quad
\label{GapEq}\,,
\end{eqnarray}
where $\Lambda$ is a momentum cut-off associated with the linearization of
the dispersion close to the CDW nodes.

For $\mu=0$, the gap equation is rather simple and reads, 
\begin{eqnarray}
\Delta_{s}(T,g,\mu=0) & = & \frac{2}{\beta}\,\cosh^{-1}\left[\cosh[\pi v_{\Delta}v_{F}\beta/g_{c}]\right.\nonumber \\
 &  & \qquad\left.\times\textrm{e}^{-\pi v_{\Delta}v_{F}\beta/g}\right],
\label{GapSol}
\end{eqnarray}
where $g_{c}=2\pi v_{\Delta}/\Lambda$ is the zero temperature critical
coupling constant. In fact, 
\begin{eqnarray}
\Delta_{s}(T=0,g,\mu=0) & = & 2\pi
v_{\Delta}v_{F}g_{c}^{-1}\left(1-\frac{g_{c}}{g}\right)\,.
\label{GapSolTZero}
\end{eqnarray}
Notice that for $g < g_c$ we find $\Delta_{s}(T=0,g<g_c,\mu=0) =0$.
Hence, the $\mu=0$ gap equation has a quantum critical point (QCP),
indicating that superconductivity occurs only above a minimal coupling
$g_{c}$. This is a general property of the nodal liquid due to the absence
of the background Fermi sea. In a Fermi liquid (where the Fermi surface
is large in comparison to all the other energy scales), the Fermi
sea is unstable to the formation of Cooper pairs between two electrons
mediated by an attractive potential, even for infinitesimal coupling 
\cite{Cooper}. In this case, the Pauli exclusion principle of the
background electrons plays the role of the interaction, making the
condensate stable even in the weak coupling limit \cite{tinkham}.
The zero temperature gap (\ref{GapSolTZero})
equals to the energy cut-off $\alpha=v_{F}\Lambda$ in the $g\to\infty$
limit.

\subsection{Zero temperature analysis}

To see how the pocket affects the QCP when $g \sim g_{c}$ we analyze
the gap equation in the zero temperature limit. At this point we 
introduce a more suitable cut-off, given by the momenta
$s_{\pm}$ that define the surfaces of constant energy in the Dirac cone,
\begin{equation}
\alpha\equiv
v_{F}^{2}\Lambda^{2}=(v_{F}s_{\pm}\pm\mu)^{2}+\Delta_{s}^{2}=const.
\label{ESurface}
\end{equation}
This new definition of the cut-off (basically replacing $\Lambda$
by $s_{\sigma}$, with $\sigma=\pm$) is convenient because it simplifies
the integration, allowing us to find simple analytical
expressions for the gap. This approximation is fairly reasonable,
since the results of the model are not to be taken literally
when $\mu$ and $\Delta_{s}$ are comparable to the energy cut-off
of the Dirac cone, $\alpha$, in which case the contribution of the 
high energy states cannot be neglected. On the other hand, 
we should be warned by the
fact that this new momentum cut-off $s_{\sigma}$ does \emph{not}
conserve the number of states of the normal phase. When calculating
thermodynamic functions, the correct cut-off is $\Lambda$, which
correctly maps the volume of the Dirac cone and avoids problems such as
loosing states in the SC phase, what would certainly have an effect
in the condensation energy. For almost all the applications,
the results are not seriously affected by the details of the cut-off
if the gap, $\Delta_{s}$, is sufficiently small in comparison to $\alpha$.

The $T=0$ gap equation becomes
\begin{eqnarray}
2\Delta_{s} & = & \frac{{g}v_{F}}{v_{\Delta}}\Delta_{s}\sum_{\sigma=\pm1}\int_{0}^{s_{\sigma}}\frac{\textrm{d}\bar{k}}{2\pi}\,\frac{\bar{k}}{E_{\mathbf{k},\sigma\mu}}\nonumber \\
 & = & \frac{{g}}{2\pi v_{F}v_{\Delta}}\Delta_{s}\left[2\alpha-2\sqrt{\Delta_{s}^{2}+\mu^{2}}\right.\nonumber \\
 &  &
 \qquad\qquad\left.-\mu\,\ln\left(\frac{\sqrt{\Delta_{s}^{2}+\mu^{2}}-\mu}{\sqrt{\Delta_{s}^{2}+\mu^{2}}+\mu}\right)\right]
 \, .
\label{gap4}
\end{eqnarray}
We rescale all the quantities by defining $x=\Delta_{s}/|\mu|$ and\[
h(g)=2\pi v_{F}v_{\Delta}\frac{g_{c}^{-1}-g^{-1}}{|\mu|}\,.\]
The $T=0$ scale invariant equation is \begin{eqnarray}
F\left(x,\, h(g)\right) & = & \sqrt{1+x^{2}}\nonumber \\
 &  & +\frac{1}{2}\,\textrm{ln}\left(\frac{\sqrt{1+x^{2}}-1}{\sqrt{1+x^{2}}+1}\right)-h(g)\nonumber \\
 & = & 0\,.\label{gap3}\end{eqnarray}

\begin{figure}
$\!\!\!\!\!\!\!\!\!\!\!\!\!\!\!\!\!\!\!\!$\includegraphics[%
  scale=0.86]{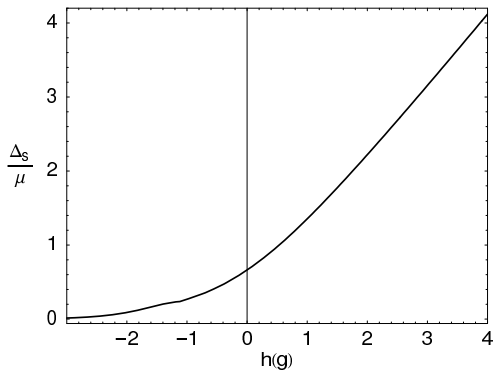}

\caption{{\small Scaling of the zero temperature gap equation versus 
the coupling constant $h(g)\propto(g_{c}^{-1}-g^{-1})/|\mu|$ .}}
\end{figure}

We see in Fig. 4  that Eq. (\ref{gap3}) has two distinct coupling
regimes: 

(\emph{i}) the \emph{strong coupling} sector $g>g_{c}$,
where the marginal physics develops, with $|\mu|\ll\Delta_{s}(0,g,\mu)$
for $g\gg g_{c}$; 

(\emph{ii}) the \emph{weak-coupling}
sector $g<g_{c}$, where the energy scale of the pocket is large in
comparison to the gap {[}\emph{i.e.} $|\mu|\gg\Delta_{s}(0,g,\mu)${]}
when $g/g_{c}\to0$. 

In the later, the system flows in the direction of
a Fermi liquid state in the weak-coupling limit ($g\ll g_{c}$), while in
the former the nodes are well defined for $g>g_{c}$ , resulting in
a nodal liquid description. We notice that the quasi-particle properties are
strongly affected by the coupling constant $g$, which separates the
marginal Fermi liquid (MFL) sector from the {}``Fermi liquid'' one, 
where the pocket
plays the role of the Fermi surface, raising the density of states
in the nodes. 

For convenience, we denote the zero temperature gap 
$\Delta_{s}(0,g,\mu)$ by $\Delta_{0\mu}$
from now on. In the strong coupling limit, $(|\mu|/\Delta_{0\mu}\ll1)$,
we may write Eq. (\ref{gap4}) as 
\[
\qquad\qquad1=\frac{{g}}{2\pi v_{\Delta}v_{F}}\left(\alpha-\Delta_{s}+\frac{\mu^{2}}{2\Delta_{s}}\right)\,,\]
whose solution is 
\begin{eqnarray}
\Delta_{0\mu} & \stackrel{g\gg g_{c}}{\longrightarrow} &
\frac{\Delta_{0}}{2}\left(1+\sqrt{1+2\mu^{2}/\Delta_{0}^{2}}\right)\,,
\label{gapMu}
\end{eqnarray}
where $\Delta_{0}\equiv\Delta(T=0,g,\mu=0)$ is given by Eq. (\ref{GapSolTZero}).
In the opposite limit, $\Delta_{0\mu}/|\mu|\ll1$, in the
weak-coupling sector, we see that Eq. (\ref{gap3}) can be expanded
in leading order in $x$, giving
\begin{eqnarray*}
F\left(x,\, h(g)\right) & \stackrel{x\to0}{\longrightarrow} &
1+\textrm{ln}\left(\frac{x}{2}\right)-h(g)
\\
 & = & 1+\textrm{ln}\left(\frac{\Delta_{s}}{2|\mu|}\right)-2\pi
 v_{F}v_{\Delta}\frac{g_{c}^{-1}-g^{-1}}{|\mu|}
\\
 & = & 0\,, 
\end{eqnarray*}
yielding, 
\begin{eqnarray}
\qquad\Delta_{0\mu} & \stackrel{g\ll g_{c}}{\longrightarrow} & 2|\mu|\,\textrm{e}^{h(g,\mu)-1}\nonumber \\
 & = & 2|\mu|\,\textrm{e}^{2\pi v_{F}v_{\Delta}(g_{c}^{-1}-g^{-1})|\mu|^{-1}-1}\,.\qquad\end{eqnarray}

\begin{figure}
\vspace{-0.05cm}$\!\!\!\!\!\!\!\!\!\!\!\!\!\!\!\!\!\!\!\!\!\!$\includegraphics[%
  scale=0.84]{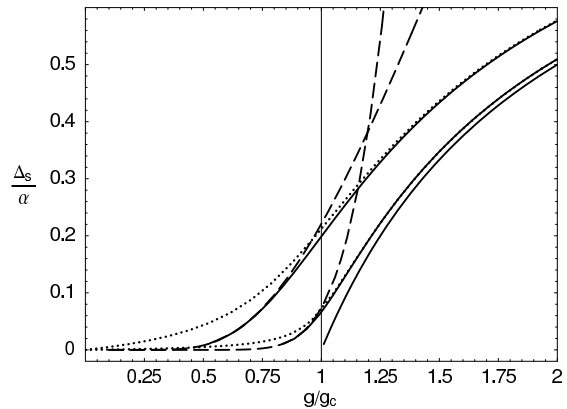}\vspace{-0.3cm}

\caption{{\small Dependence of the zero temperature gap (normalized by the
cut-off $\alpha$) with the coupling constant $g$.  Solid lines: numeric solution
of the gap equation (\ref{gap3}); dotted: strong coupling approximation
($|\mu|\ll\Delta_{0\mu}$); dashed: weak coupling one ($|\mu|\gg\Delta_{0\mu}$). 
We have set $|\mu|/\alpha=0,\,0.1$
e $0.3$ from the bottom to the top.
Notice that the QCP at $\mu=0$ is suppressed by the pocket formation ($|\mu|>0$).}}
\end{figure}

Although the strong coupling approximation is rigorously valid only for
$g\gg g_{c}$, and the weak coupling one for $g\ll g_{c}$, these two
approximations are remarkably good in almost the entire coupling range
of their respective sectors (as shown in Fig. 5) provided that $|\mu|/\alpha$
is small. However, to find sensible results, one should consider that
the valid coupling range of the theory is limited not too far above
the critical coupling $g_{c}$, 
in order to keep the ratio $\Delta_{0\mu}/\alpha$ small (see Fig. 5).

\subsection{Finite temperatures}

Let us return to Eq. (\ref{GapEq}). After some algebraic manipulation
(see the details in appendix A), the gap equation in the strong coupling
regime assumes the form

\begin{eqnarray}
 &  & \cosh\!\left(\beta\Delta_{s}/2\right)\textrm{e}^{-\mu^{2}\beta\,\tanh(\beta\Delta_{s}/2)/(4\Delta_{s})}\nonumber \\
 &  & \qquad\qquad=\,\,\cosh\!\left(\pi
 v_{\Delta}v_{F}\beta/g_{c}\right)\textrm{e}^{-\pi
 v_{\Delta}v_{F}\beta/g}\,.\qquad
\label{gapSCoupling}
\end{eqnarray}
The quantity $\tanh(\beta\Delta_{s}/2)/\Delta_{s}$ changes very little
with $\beta$ in the whole temperature interval. In a first approximation,
we can obtain the analytical expression of the low temperature gap
by replacing the gap inside the exponential by its zero temperature
value $\Delta_{0\mu}$. This substitution leads to:
\begin{eqnarray*}
\Delta_{s}(T,g,\mu) & \sim & \frac{2}{\beta}\,\cosh^{-1}\left[\cosh\left(\pi
    v_{\Delta}v_{F}\beta/g_{c}\right)\right.
\\
 &  & \left.\times\,\textrm{e}^{-\pi v_{\Delta}v_{F}\beta/g}
\textrm{e}^{\mu^{2}\beta\,\tanh(\beta\Delta_{0\mu}/2)/(4\Delta_{0\mu})}\right],
\end{eqnarray*}
valid in strong coupling for small $\mu/\alpha$. Close to the phase
transition, Eq. (\ref{GapEq}) gives 
\begin{eqnarray}
\Delta_{s}(T) & \stackrel{T\to T_{c}}{\longrightarrow} & \left\{ 
\begin{array}{cl}
2\sqrt{\frac{\Delta_{0}}{\beta_{c}}+\frac{\mu^{2}}{2}}\, t^{\frac{1}{2}} &
\,,|\mu|/\Delta_{0\mu}\ll1
\\
\\
\frac{1}{\beta_{c}}\left[\frac{7\zeta(3)}{8\pi^{2}}+\frac{1}{2\beta_{c}^{2}\mu^{2}}\right]^{-\frac{1}{2}}t^{\frac{1}{2}}
& \,,|\mu|/\Delta_{0\mu}\gg1\,,
\end{array}\right.
\label{GapTc2}
\end{eqnarray}
where $t\equiv(T_{c}-T)/T_{c}$ is the reduced temperature and $\zeta(x)$
is the Zeta function. The critical temperature is also calculated
from the gap equation, (\ref{GapEq}), in the $\Delta_{s}\to0$ limit, giving
\begin{equation}
T_{c}=\left\{ 
\begin{array}{cl}
\frac{1}{2k_{B}\ln4}\left[\Delta_{0}+\sqrt{\Delta_{0}^{2}+\mu^{2}\ln4}\right]
& \,,|\mu|/\Delta_{0\mu}\ll1
\\
\\\frac{|\mu|\gamma}{k_{B}\pi}\,\textrm{e}^{\alpha(1-g_{c}/g)|\mu|^{-1}-1} &
\,,|\mu|/\Delta_{0\mu}\gg1\,,
\end{array}\right.
\label{TcMu}
\end{equation}
where $\ln\gamma\sim0.577$ is the Euler constant. In the particle-hole symmetric
case ($\mu=0$), we have $T_{c}=\Delta_{0}/(k_B\ln4)$ and $\Delta_{s}(T\to
T_{c},g,0)=2\Delta_{0}t^{\frac{1}{2}}/\ln2$ (see appendix A for details). 

We see that the existence of a pocket suppresses the QCP ($T=0$) separating 
the normal and SC phases (see Fig. 5). 
This effect is due to the establishment of
the background Fermi sea, which stabilizes the Cooper pairs for
an arbitrarily small coupling. The thermal effect on the gap recovers
the parametric phase transition with the coupling constant $g$, as
displayed in Fig. 6 (top) by noting the presence of a minimal coupling
(say, $g_{0}(T,\mu)$, with $g_{0}(0,0)=g_{c}$), below which
$\Delta_{s}(g<g_{0},\mu)=0$. The explanation can be found in the strong 
dependence of the critical
temperature $T_{c}$ with $g$, as shown in Fig. 6 (bottom). At a
given non-zero temperature $T$, a minimal coupling is required to satisfy
$T_{c}$($g>g_{0}$) $>T$.

\begin{figure}
$\!\!\!\!\!\!\!\!\!\!\!\!\!\!\!\!\!\!\!\!\!$\includegraphics[%
  scale=0.817]{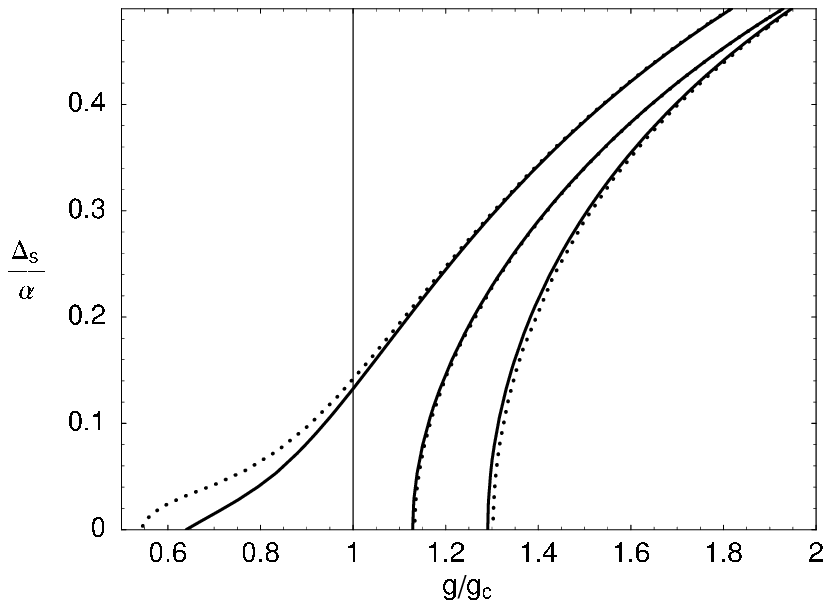}\\
\vspace{-0.5cm}

$\!\!\!\!\!\!\!\!\!\!\!\!\!\!\!\!\!\!\!\!\!\!\!\!\!\!$\includegraphics[%
  scale=0.81]{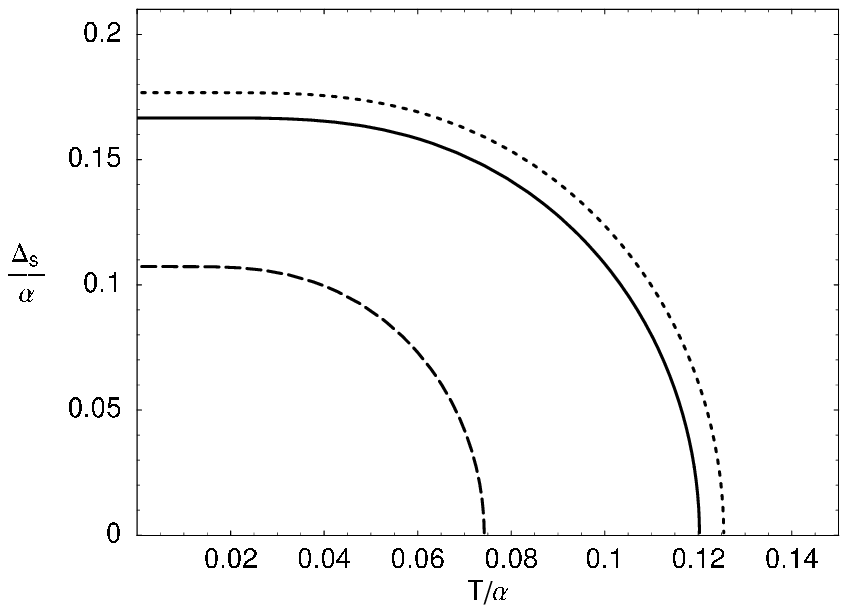}

\caption{{\small Top: SC gap $\Delta_{s}(T,g,\mu)$ vs. the coupling constant
$g/g_{c}$. Solid lines: numeric solution of the gap equation (\ref{GapEq});
dotted lines: strong coupling solution (analytic). From left to right:
$k_{B}T/\alpha=0.005,\,0.1$, 0.2 and $|\mu|/\alpha=0.2,\,0.2,\,0.1$,
respectively. Bottom: SC gap vs. temperature. The scales are normalized
by the energy cut-off $\alpha$ of the Dirac cone. Dotted line: ($|\mu|/\alpha=0.06,\, g/g_{c}=1.2$);
solid: ($\mu/\alpha=0,\, g/g_{c}=1.2$); dashed: ($|\mu|/\alpha=0.06,\, g/g_{c}=1.1$).}}
\end{figure}

\section{Thermodynamics}

In this section, we calculate the thermodynamic functions starting
from the partition function $Z$ of the nodal fermions. The partition function 
is defined
as usual from the original Hamiltonian (\ref{H1}), written in a diagonal
basis of eigenstates indexed by $\mathbf{k}$, $\gamma=\pm1$ (for the two particle-hole branches), $\sigma =\pm1$, 
and with eigenvalues $E^a_\mathbf{k}=\pm E_{\mathbf{k},\sigma\mu}$:
\begin{eqnarray*}
Z\,=\,\textrm{e}^{-\beta\Omega} & = & \textrm{tr}\,\textrm{e}^{-\beta H}
\\
 & = & \textrm{e}^{-\beta
   g^{-1}\Delta_{s}^{2}}\,\prod_{\mathbf{k},a}\,\sum_{n_{\mathbf{k}}^{a}=0}^{1}\,\langle n_{\mathbf{k}}^{a}|\textrm{e}^{-\beta\, E_{\mathbf{k}}^{a}\, n_{\mathbf{k}}^{a}}|n_{\mathbf{k}}^{a}\rangle
\\
 & = & \textrm{e}^{-\beta
   g^{-1}\Delta_{s}^{2}}\,\prod_{\mathbf{k},\gamma,\sigma}\,\left(1+\textrm{e}^{-\beta\,\gamma E_{\mathbf{k},\sigma\mu}}\right)\,,
\end{eqnarray*}
where $\Omega$ is the thermodynamic potential. The Hamiltonian
includes the term $\Delta_{s}^{2}/g$, in order to
give the correct condensation energy.
The thermodynamic potential, $\Omega=\Omega_{0}+\Delta_{s}^{2}/g$,
is given by 
\begin{eqnarray}
\Omega_{0}(T) & = &
-\frac{1}{\beta}\sum_{\mathbf{k},\sigma}\textrm{ln}\left[2+2\textrm{cosh}(\beta{E}_{\mathbf{k},\sigma\mu})\right]\nonumber 
\\
 & = & -\frac{v_{F}}{\pi\beta
   v_{\Delta}}\sum_{\sigma}\int_{0}^{\Lambda}\textrm{d}\bar{k}\,\bar{k}\,\textrm{ln}\left[2+2\textrm{cosh}(\beta E_{\mathbf{k},\sigma\mu})\right].
\nonumber 
\\
\label{thermoPot}
\end{eqnarray}
If $\Omega_{n}$ is the thermodynamic potential in the normal phase, the 
condensation energy, $\Omega_{n}(0)-\Omega_{0}(0)= H_{c}^{2}(0)/(8\pi)$,
is given in terms of the zero temperature critical field, $H_{c}$,
shown in Fig. 7. Analogously, the internal energy, $E=E_{0}+\Delta_{s}^{2}/g$,
is given by
\begin{eqnarray}
E_{0}(T) & = & \sum_{\mathbf{k},\gamma,\sigma}\,\gamma
E_{\mathbf{k},\sigma\mu} n_{\mathbf{k}}^{\gamma}\nonumber 
\\
 & = & -\frac{v_{F}}{2\pi
   v_{\Delta}}\sum_{\sigma}\int_{0}^{\Lambda}\textrm{d}\bar{k}\,\bar{k\,}E_{\mathbf{k},\sigma\mu}\tanh\left(\frac{\beta E_{\mathbf{k},\sigma\mu}}{2}\right).\quad\nonumber 
\\
\label{intEnergy}
\end{eqnarray}
where $n_{\mathbf{k}}^{\gamma}=(\textrm{e}^{\gamma\beta E_{\mathbf{k},\sigma\mu}}+1)^{-1}$
is the Fermi-Dirac distribution indexed by $\sigma,\gamma=\pm1$. 

\begin{figure}[]


$\!\!\!\!\!\!\!\!\!\!\!\!\!\!\!\!\!\!\!\!\!$\includegraphics[width=8.1cm]{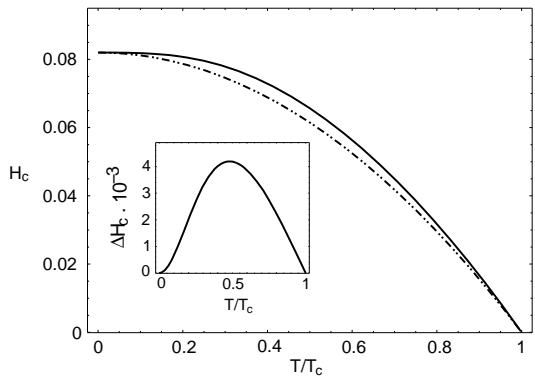}




\caption{{\small Solid line: critical field  $H_c$ dependence with
    temperature, in units of $ \alpha/\sqrt{g_c} $ for $\mu=0$ and
    $g/g_c=1.1$; dotted: 
empirical law  $H_c(0)[1 -T^2/T_c^2]$. The difference  between the two curves is shown in the inset.}}
\end{figure}

According to the standard thermodynamic relations, the specific heat
is defined by 
\begin{equation}
C_{V}=T\frac{\textrm{d}S}{\textrm{d}T}=-\beta\frac{\textrm{d}S}{\textrm{d}\beta}\,,
\label{Cv}
\end{equation}
where $S=(E-\Omega)/T=-\left(\frac{\partial\Omega}{\partial T}\right)_{V}$
is the electronic entropy due to the Dirac fermions. At low temperature,
the gap is practically independent on the temperature. It is easy 
to check that the specific heat dependence with temperature
in this limit for $\mu=0$ is given by:
\begin{eqnarray*}
C_{V} & \stackrel{T\ll T_{c}}{\longrightarrow} & \frac{1}{2\pi
  v_{F}v_{\Delta}}\,\int_{\Delta_{s}}^{E_{\Lambda}}\textrm{d}E\,
E^{3}\,\textrm{sech}^{2}\!\left(\frac{\beta E}{2}\right)
\\
 & \stackrel{\Lambda\to\infty}{\longrightarrow} & \frac{6k_{B}}{\pi
  v_{F}v_{\Delta}}\Delta_{s}^{2}\,\textrm{e}^{-\beta\Delta_{s}}\,,
\end{eqnarray*}
leading to the expected exponential behavior of $C_V$ with the gap. 

A more interesting result is related to the jump of the specific heat in 
the normal-SC phase transition.
The calculation is given in appendix B for the weak and strong coupling
regimes. It results in two well defined limits: the marginal one
($|\mu|\beta_{c}\ll1$), 
\[
\left.\frac{\Delta
    C_{V}}{C_{n,V}}\right|_{T_{c}}=\frac{2\ln4}{9\zeta(3)}\,\left(\ln4+\frac{\beta_{c}^{2}\mu^{2}}{2}\right)\,\geq\,0.35,
\]
where the equality holds for $\mu=0$; and the Fermi liquid limit
\begin{eqnarray}
\left.\frac{\Delta C_{V}}{C_{n,V}}\right|_{T_{c}} & = &
\frac{3}{2\pi^{2}}\frac{1}{\frac{7\zeta(3)}{8\pi^{2}}+\frac{1}{2\beta_{c}^{2}\mu^{2}}}\,\leq\,1.43\,,
\label{DeltaCweak}
\end{eqnarray}
which recovers the BCS result for $\beta_{c}|\mu|\gg1$. 

The jump observed in NbSe$_{2}$ \cite{craven,Kobayashi,garoche} 
($\Delta C/C_{n}\sim2$) is a good indication in favor of a conventional
Fermi liquid and BCS behavior. In TaSe$_{2}$,
however, where the transport is marginal and the quasi-particles are
not well defined in the Landau sense \cite{dordevic3} ($\tau\omega<1$,
where $\tau^{-1}$ is the scattering rate), the picture can be very
different. In the nodal liquid case, 
the specific heat jump is strongly attenuated
due to the vanishing density of states (DOS) in the Fermi surface,
resulting in the universal constant $\Delta C_{V}/C_{n}$ $=0.35$.
The plot of the specific heat displayed in Fig. 8 shows that the temperature
dependence of the normal CDW phase is quadratic. 
As the DOS is raised by a pocket around the
nodes, the jump grows in direction to the BCS value of $1.43$,
which corresponds to the weak coupling limit. However, we notice that
the nodes cease to be well defined in the presence of large pockets.
In this case, the pairing ansatz adopted in sec. II and the role of
piezoelectricity in the electron-phonon coupling are questionable.

\begin{figure}[]
$\!\!\!\!\!\!\!\!\!\!\!\!\!\!\!\!\!\!\! $\includegraphics[%
  scale=0.81]{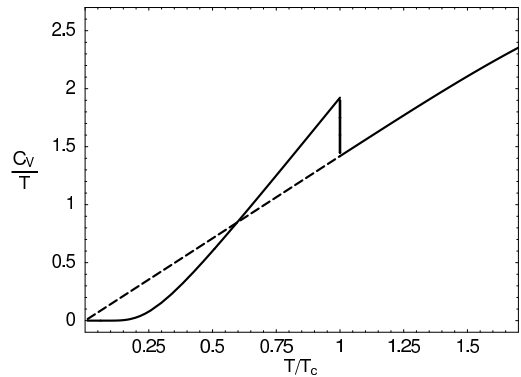} \qquad

\caption{{\small Specific heat $C_{V}/T$ vs. temperature for $\mu=0$, in 
units of $k_B^2/g_c$.
The jump occurs at $k_{B}T_{c}=\Delta_{0}/\ln4$. Dashed line: normal
behavior in the absence of the SC gap.}}
\end{figure}

\section{Coherence factors}

In this section we calculate two basic properties of the superconductor:
the acoustic attenuation and the nuclear spin relaxation rate in the
absence of impurities.

\subsection{Acoustic attenuation }

The ultra-sound attenuation results from the resonant absorption of
the longitudinal phonons in the solid \cite{schrieffer}. The absorption
rate is proportional to the imaginary part of the charge susceptibility
\cite{Mahan}:
\begin{equation}
\alpha_{s}(\mathbf{q})=-\lambda_{1}^{2}\,\lim_{\omega\to0}\left[\frac{1}{\omega}\textrm{Im}\,\chi^{c}(\mathbf{q},\omega)\right]
\label{ultraSound}
\end{equation}
in the $q\to0$ limit, since the phonon wavelength is much larger
than the typical electronic wavelength. This property is connected
to the superconductor \emph{coherence factors}, which basically define
the probability amplitude of quasi-particle transitions between two
states represented by the pairs space $(\mathbf{k}\uparrow,-\mathbf{k}\downarrow)$
\cite{schrieffer,tinkham}. These factors conserve the time reversal
symmetries of the interaction involved in the transition. They are
usually divided into type I, for interactions which preserve the time-reversal
symmetry (like in the electron-phonon coupling) and type II when this
symmetry is broken, like in the spin exchange interaction. The charge
susceptibility is defined in terms of the time ordered charge density
correlation function. All the correlation functions used in this article
are defined in appendix C. Using the spinor defined in (\ref{spinor}), 
the charge density operator is given by: 
\begin{eqnarray}
\rho(\mathbf{q}) & = &
\sum_{\mathbf{k},\sigma,a}\psi_{a\,\sigma}^{\dagger}(\mathbf{k}-\mathbf{q}/2)\,\psi_{a\,\sigma}(\mathbf{k}+\mathbf{q}/2)\nonumber 
\\
 & = &
 \sum_{\mathbf{k}}\,\Psi^{\dagger}(\mathbf{k}-\mathbf{q}/2)\,\tau_{3}\eta_{0}\Psi(\mathbf{k}+\mathbf{q}/2) \, .
\label{rhoDens}
\end{eqnarray}
We define: $\stackrel{\leftrightarrow}{G}_{+}\equiv\stackrel{\leftrightarrow}{G}(\mathbf{k}+\mathbf{q}/2,i\omega_{n}+i\omega)$
and
$\stackrel{\leftrightarrow}{G}_{-}\equiv\stackrel{\leftrightarrow}{G}(\mathbf{k}-\mathbf{q}/2,i\omega_{n})$,
so that the electronic charge susceptibility reads
\begin{equation}
\chi^{c}(\mathbf{q},i\omega)=\frac{1}{\beta}\textrm{Tr}\sum_{\mathbf{k},\omega_{n}}\stackrel{\leftrightarrow}{G}_{+}\tau_{3}\eta_{0}\,\stackrel{\leftrightarrow}{G}_{-}\tau_{3}\eta_{0}\,.
\label{ChiChargeGF}
\end{equation}
It is convenient to define the gapless Dirac fermions dispersion
by $\epsilon_{\mathbf{k}}=v_{F}\bar{k}$, and the quantity $\epsilon_{o}=\sqrt{v_{F}^{2}(\bar{k}^{2}+\bar{q}^{2}}$).
After evaluating the trace and the sum over the fermionic Matsubara
frequencies, the imaginary part of the susceptibility reads:
\begin{eqnarray}
\textrm{Im}\,\chi^{c}(\mathbf{q},\omega\to0) & = &
\frac{\omega}{\epsilon_{\mathbf{q}/2}}\frac{v_{F}}{\pi
  v_{\Delta}}\,\int_{0}^{\Lambda}\textrm{d}\bar{k}\,\bar{k}\sum_{\sigma=\pm1}\frac{\partial n(E_{o,\sigma\mu})}{\partial E_{o,\sigma\mu}}
\nonumber 
\\
&  &
\qquad\times\frac{\epsilon_{o}+\sigma\mu}{\epsilon_{o}E_{o,\sigma\mu}}\sqrt{\epsilon_{o}^{2}-\epsilon_{\mathbf{q}/2}^{2}},
\label{ImXi}
\end{eqnarray}
where $E_{o,\sigma\mu}=\sqrt{(\epsilon_{o}+\sigma\mu)^{2}+\Delta_{s}^{2}}$,
and $n$ is the Fermi-Dirac distribution. Replacing Eq. (\ref{ImXi})
into Eq. (\ref{ultraSound}), we obtain the ultra-sound attenuation rate:
\begin{eqnarray*}
\alpha_{s} & \stackrel{q\to0}{\longrightarrow} &
-\frac{1}{\epsilon_{\mathbf{q}/2}}\frac{\lambda_{1}^{2}}{\pi
  v_{\Delta}v_{F}}\,\sum_{\sigma=\pm1}\int_{0}^{\alpha}\frac{\textrm{d}\epsilon\,\epsilon}{E_{\sigma\mu}}\,(\epsilon+\sigma\mu)\frac{\partial n(E_{\sigma\mu})}{\partial E_{\sigma\mu}}.
\end{eqnarray*}
The temperature dependence of $\alpha_s$ 
is displayed in Fig. 9 and shows a power-law
behavior near the phase transition. This result
is compared with the BCS curve 
$\alpha_{s}/\alpha_{n}=2/(\textrm{e}^{\beta\Delta_{s}}+1)$ \cite{tinkham}.

\begin{figure}[]
\vspace{0.cm}$\!\!\!\!\!\!\!\!\!\!\!\!\!\!$\includegraphics[%
  scale=0.81]{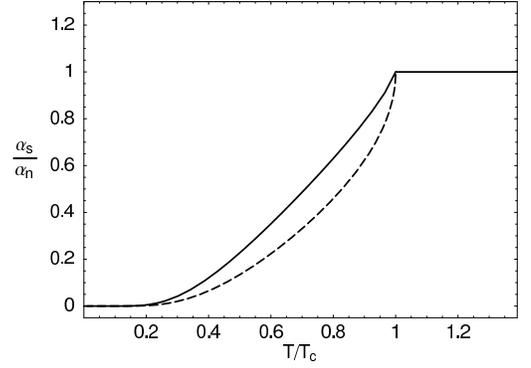}

\vspace{-0.2cm}

\caption{{\small Temperature dependence of the acoustic attenuation rate normalized
by the normal phase rate. Solid: our model ($\mu=0$ and $g/g_c=1.1$); dashed: BCS model. }}
\end{figure}

\subsection{NMR relaxation}

The NMR relaxation has its origin on the hyperfine interaction between the
nuclear spins and the electrons. The relaxation rate measures
the nuclear spin time-variation along an arbitrary direction of the
spin space, say $\hat{\mathbf{b}}$. The condensate exhibits no paramagnetism
in the singlet channel, where the total spin of the pairs is zero.
Since the Zeeman and hyperfine energies are usually small in comparison
to the gap, the only processes that contribute to the spin relaxation
are thermally excited quasi-particles. The inverse of the spin relaxation
is proportional to the local magnetic susceptibility projected along
$\hat{\mathbf{b}}$, 
\begin{equation}
T_{1}^{-1}(\hat{\mathbf{b}})=-\lambda_{2}^{2}\,\sum_{\mathbf{q}}\,\lim_{\omega\to0}\left[\frac{1}{\omega}\textrm{Im}\,\chi_{\hat{\mathbf{b}}}^{s}(\mathbf{q},\omega)\right],
\label{NMR}
\end{equation}
where $\chi_{\hat{\mathbf{b}}}^{s}(\omega)$ is given in terms of
the normal directions of the spin space by: 
$\chi_{\hat{\mathbf{b}}}^{s}(\omega)=\sum_{ij}\,(\delta^{ij}-b^{i}b^{j})\chi_{ij}^{s}(\omega)\,,$with
$i,j=1,2,3$ representing the $x,\, y,\, z$ directions, respectively
(see appendix C). 

Before defining the spin density operator, we must introduce the spin
degrees of freedom in the spinor representation, Eq. (\ref{spinor}). 
This is naturally done in the Balian-Werthamer (BW) space \cite{balian},
\begin{equation}
\Psi_{a}(\mathbf{k})=\left(\begin{array}{c}
\zeta_{a}(\mathbf{k})
\\
-i\sigma_{2}\,\zeta_{a}^{\dagger}(-\mathbf{k})
\end{array}\right)=\left(\begin{array}{c}
\psi_{a\uparrow}(\mathbf{k})
\\
\psi_{a\downarrow}(\mathbf{k})
\\
-\psi_{a\downarrow}^{\dagger}(-\mathbf{k})
\\
\psi_{a\uparrow}^{\dagger}(-\mathbf{k})
\end{array}\right),
\label{BWS}
\end{equation}
which contains an additional spin subspace 
\[
\zeta(\mathbf{k})=\left(\begin{array}{c}
\psi_{\uparrow}(\mathbf{k})
\\
\psi_{\downarrow}(\mathbf{k})\end{array}\right),
\]
 inside the regular Nambu space, $(\uparrow\mathbf{k},\downarrow-\mathbf{k})$.
We have defined a new set of Pauli matrices $\sigma_{\mu}=(\sigma_{0},\vec{\sigma})$
which operate in this new space. The general spin density operator
is 
\begin{eqnarray}
S_{i}(\mathbf{q}) & = &
\frac{1}{2}\sum_{\mathbf{k}\sigma\sigma^{\prime}a}\,\psi_{a,\sigma}^{\dagger}(\mathbf{k}-\mathbf{q}/2)\,\sigma_{i}^{\sigma\,\sigma^{\prime}}\psi_{a,\sigma^{\prime}}(\mathbf{k}+\mathbf{q}/2)\nonumber 
\\
 & = &
 \frac{1}{2}\sum_{\mathbf{k}\in\frac{1}{2}\textrm{B.Z.}}\Psi^{\dagger}(\mathbf{k}-\mathbf{q}/2)\,\sigma_{i}\tau_{0}\eta_{0}\,\Psi(\mathbf{k}+\mathbf{q}/2)\,,
\qquad
\label{S}
\end{eqnarray}
where $i=1,2,3$ are the spin directions, $\mathbf{k}$ is summed
in half Brillouin zone, and $\sigma,\sigma^{\prime}$=$\uparrow\downarrow$
are the spin indexes. It is not difficult to check that the Hamiltonian
(\ref{H1}), written in the BW space, is given by (see appendix D)
\begin{eqnarray}
H & = &
\sum_{\mathbf{k}\in\frac{1}{2}\textrm{B.Z.}}\Psi^{\dagger}(\mathbf{k})\left[v_{F}\,\sigma_{0}\tau_{0}\vec{\eta}\cdot\bar{\mathbf{k}}-\Delta_{s\,}\sigma_{3}\tau_{1}\eta_{2}\right.\nonumber 
\\
 &  &
 \qquad\qquad\qquad\left.-\mu\,\sigma_{0}\tau_{3}\eta_{0}\right]\Psi(\mathbf{k})\,.
\label{HBW}
\end{eqnarray}
The matrix inside the parenthesis defines the new dispersion tensor
$\stackrel{\leftrightarrow}{\omega}_{\mathbf{k}}$ for the Green function
(\ref{GreenF1}), $\stackrel{\leftrightarrow}{G}\,=(i\omega_{n}-\stackrel{\leftrightarrow}{\omega})^{-1}$.
Notice that the BW Green function is very similar to the previous one, 
except for the size of the Hamiltonian space, which now is 8$\times$8.

The pairing term brings something new, because of the broken time-reversal
symmetry of the SC phase, expressed by the anti-symmetric property
of the Pauli matrix $\eta_{2}$ under the transposition: $\eta_{\alpha}^{a\, b}\to\eta_{\alpha}^{b\, a}$.
 We will soon explore the physical consequences of this broken symmetry. From Eq. (\ref{PiS}) 
the spin susceptibility tensor is given by
\begin{equation}
\chi_{ij}^{s}(\mathbf{q},i\omega)=\frac{1}{4\beta}\,\textrm{Tr}\!\!\!\sum_{\mathbf{k}\in\frac{1}{2}\textrm{B.Z.}}\sum_{\omega_{n}}\stackrel{\leftrightarrow}{G}_{+}\sigma_{i}\tau_{0}\eta_{0}\,\stackrel{\leftrightarrow}{G}_{-}\sigma_{j}\tau_{0}\eta_{0}\,.
\label{SpinCorr}
\end{equation}
Notice that the product $\sigma_{i}\tau_{0}\eta_{0}\stackrel{{\leftrightarrow}}{G}\sigma_{i}\tau_{0}\eta_{0}\,=\,\stackrel{{\leftrightarrow}}{G}$
for $i=3$. For $i=1,2$, the anti-commutative matrices $\eta_{i}$
lead to a sign change in the gap term of $\stackrel{\leftrightarrow}{\omega}$
inside the Green function, implying $\Delta_{s}\to-\Delta_{s}$. Thus, 
the $i=1,\,2$ (\emph{i.e.} $x,\, y$) directions have the same
coherence factors of the charge susceptibility, 
\begin{eqnarray}
\chi_{xx}^{s}(\mathbf{q},\omega) & \!\!=\!\! &
\chi_{yy}^{s}(\mathbf{q},\omega)=\frac{1}{4}\chi^{c}(\mathbf{q},\omega)\,.
\label{chiii}
\end{eqnarray}
This property is better illustrated in the $\mu=0$ case, where
\begin{eqnarray*}
\chi_{xx}^{s} & = &
\frac{1}{\beta}\sum_{\mathbf{k},\omega_{n}}\frac{(\vec{\epsilon}_{-}\cdot\vec{\epsilon}_{+}-\Delta_{s}^{2})-\omega_{n}(\omega_{n}+\omega)}{\left[\omega_{n}^{2}+E_{-}^{2}\right]\left[(\omega_{n}+\omega)^{2}+E_{+}^{2}\right]}=\chi_{yy}^{s},
\end{eqnarray*}
and
\begin{eqnarray*}
\chi_{zz}^{s} & = &
\frac{1}{\beta}\sum_{\mathbf{k},\omega_{n}}\frac{(\vec{\epsilon}_{-}\cdot\vec{\epsilon}_{+}+\Delta_{s}^{2})-\omega_{n}(\omega_{n}+\omega)}{\left[\omega_{n}^{2}+E_{-}^{2}\right]\left[(\omega_{n}+\omega)^{2}+E_{+}^{2}\right]},
\end{eqnarray*}
with $\vec{\epsilon}_\mathbf{k}\equiv v_F\bar{\mathbf{k}}$, and the indexes $\pm$ 
representing the momentum, $\pm\to\mathbf{k}\pm\mathbf{q}/2$.
Notice the sign difference in front of $\Delta_{s}^{2}$ between the
$zz$ and the other two components. This difference 
gives rise to an axial anisotropy
in the $z$ direction of the spin space. This anisotropy is a consequence
of the broken time-reversal symmetry induced by the finite momentum
of the pairs, $\mathbf{Q}_{i}$, which defines the CDW wave vectors. 
This broken symmetry is 
reflected in the appearance of a spin structure oriented in the $z$
direction (indicated by the $\sigma_{3}$ matrix)
in the pairing term of the BW Hamiltonian (\ref{HBW}). Therefore, 
we conclude that the $z$ direction in the spin space corresponds
to the CDW direction $\mathbf{Q}_{i}$ in the $k$-space, since it
is the only rotational symmetry broken in the crystal. The calculation
of the imaginary part of the $\chi_{zz}^{s}$ susceptibility reads:
\begin{eqnarray}
\textrm{Im}\,\chi_{zz}^{s}(\mathbf{q},\omega\to0)\!\! & = &
\!\!\frac{\omega}{4\pi}\frac{v_{F}}{v_{\Delta}}\,\int_{0}^{\Lambda}\textrm{d}\bar{k}\,\bar{k}\nonumber 
\\
 &  & \!\!\left\{ \frac{\partial
     n(\tilde{E})}{\partial\tilde{E}}\,\frac{2\Delta_{s}^{2}}{\tilde{\epsilon}_{\mathbf{k}}\tilde{E}}\frac{\epsilon_{\mathbf{q}/2}^{2}-\mu^{2}}{\tilde{\epsilon}_{\mathbf{k}}^{2}-\mu^{2}}\right.\nonumber 
\\
 &  &
 \!\!\times\textrm{Re}\!\left[\frac{\theta\left(\tilde{\epsilon}_{\mathbf{k}}-|\mu|\right)\tilde{\epsilon}_{\mathbf{k}}+\theta\left(|\mu|-\tilde{\epsilon}_{\mathbf{k}}\right)|\mu|}{\sqrt{(\epsilon_{\mathbf{q}/2}^{2}-\mu^{2})(\tilde{\epsilon}_{\mathbf{k}}^{2}-\epsilon_{\mathbf{q}/2}^{2})}}\right]\nonumber 
\\
 &  & \!\!+\sum_{\sigma=\pm1}\,\frac{\partial n(E_{o,\sigma\mu})}{\partial
   E_{o,\sigma\mu}}\,\frac{E_{o,\sigma\mu}}{(\epsilon_{o}+\sigma\mu)}\nonumber 
\\
 &  &
 \qquad\quad\left.\times\frac{\sqrt{\epsilon_{o}^{2}-\epsilon_{\mathbf{q}/2}^{2}}}{\epsilon_{\mathbf{q}/2}\epsilon_{o}}\right\} 
\end{eqnarray}
where 
\[
\tilde{\epsilon}_{\mathbf{k}}=\textrm{Re}\sqrt{\epsilon_{\mathbf{k}}^{2}+\epsilon_{\mathbf{q}/2}^{2}-\mu^{2}},
\]
$\tilde{E}_{\mathbf{k}}=\sqrt{\tilde{\epsilon}_{\mathbf{k}}^{2}+\Delta_{s}^{2}}$ and $\epsilon_{o}$ follows the definition of the previous subsection.
The $\chi^s_{xx,yy}$ can be obtained from the substitution of Eq.
(\ref{ImXi}) into Eq. (\ref{chiii}). Noting that $\chi^s_{xy}=\chi^s_{xz}=\chi^s_{yz}=0$, the NMR relaxation
rate along a given direction $\hat{\mathbf{b}}$  gives
\begin{equation}
\frac{1}{T_{1}}(\hat{\mathbf{b}})=\lambda_{2}^{2}\,\int_{0}^{\Lambda}\frac{\textrm{d}\bar{q}}{2\pi}\,\bar{q}\lim_{\omega\to0}\left[\frac{1}{\omega}\sum_i\left( b_i^2-1\right) \textrm{Im}\,\chi_{ii}^{s}\right]\,.\label{Proj}\end{equation}

\begin{figure}
$ \!\!\!\!\!\!\!\!\!\! $\includegraphics[%
  scale=0.86]{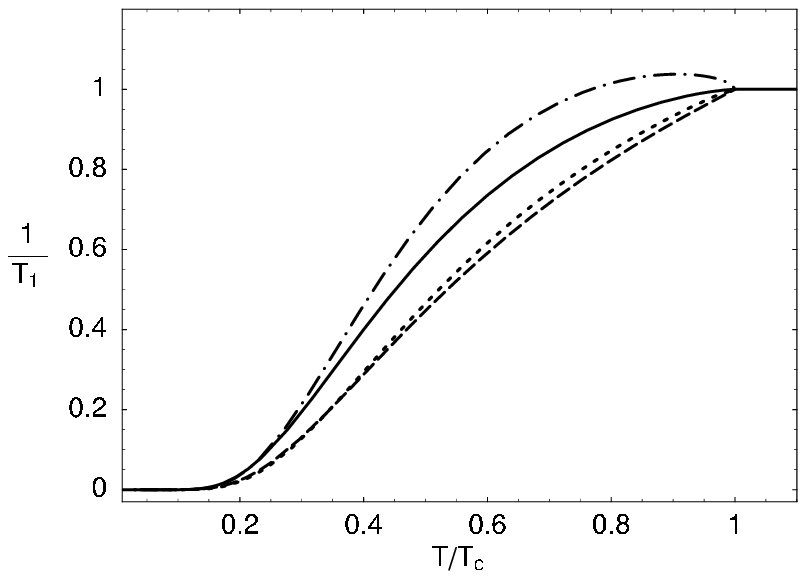}$\qquad\qquad\qquad$

\caption{{\small Temperature dependence of the NMR relaxation rate normalized
to the normal phase relaxation for $g/g_c=1.1$. Dashed ($\mu=0$) and dotted ($|\mu|/\alpha=0.05$)
lines: NMR response along the in-plane directions [$\varphi=\frac{\pi}{2}$ case of Eq. (\ref{Proj2})]; 
solid ($\mu=0$) and dot-dashed ($|\mu|/\alpha=0.05$) lines:
NMR response along the normal $c$ axis ($\varphi = 0$). The pocket produces
a small Hebel-Slichter peak, indicated by the dot-dashed line.}}
\end{figure}

In Fig. 10 we distinguish the two anisotropic principal directions, 
for in-plane  $\hat{\mathbf{b}}$ vectors and out-of-plane ones directed along the normal $c$ axis.
 A small
Hebel-Slichter peak is formed for finite $\mu$, 
but no peak is observed
for $\mu=0$. 

The $zz$ component of the susceptibility carries coherence factors
with the symmetry of the spin interactions (\emph{i.e.} they are odd
by interchanging $\mathbf{k}$ $\to-\mathbf{k}$), while the $xx$
and $yy$ components are analogous to the charge susceptibility {[}see
Eq. (\ref{chiii}){]}. This is easily understood by a qualitative
argument with the aid of Eq. (\ref{Proj}). Consider the
CDW direction $\mathbf{Q}_{1}$. The $\mathbf{Q}_{1}$
direction (or equivalently the $\mathbf{b}=\hat{z}$ direction for
the spin, according to our previous discussion) affects the electronic
spin correlations in the normal directions to $\mathbf{Q}_{1}$, meaning
the $xy$ plane. The NMR direction $\mathbf{b}=\hat{z}$ is affected
by the susceptibility components $\chi_{xx}$ and $\chi_{yy}$ but
not by the $\chi_{zz}$ one {[}see Eq. (\ref{Proj}){]}. The CDW introduces
an additional time-reversal broken symmetry to the spin correlations
in the $\mathbf{Q}_{1}$ ($xy$) plane, explaining why the related
coherence factors have the same symmetry of the charge interactions. 
On the other hand, the planes which
are normal to the $\mathbf{Q}_{1}$ plane are affected by the $\chi_{zz}$
component, which conserves the odd symmetry of the spin interactions.
In summary, the NMR relaxation in the $\hat{\mathbf{b}}=\mathbf{Q}_{1}$ 
direction (in $k$-space) is therefore associated to a \emph{charge-like} 
symmetry, like in the phonon attenuation response, while the NMR directions
which are normal to $\mathbf{Q}_{1}$ have a mixed symmetry and exhibit
a more intense response. The same analysis
applies to the $\mathbf{Q}_{2,3}$ vectors separately. The NMR pattern
in the $k$-space results from the superposition
of the contributions due to each vector $\mathbf{Q}_{i}$ ($i=1,2,3$)
of the triple-CDW. As  each vector $\mathbf{Q}_{i}$ is rotated with
respect to the other two by $\frac{2\pi}{3}$ and $\frac{4\pi}{3}$ (see
Fig 1), if we define  the contribution of each 
CDW direction to the NMR response
along an arbitrary direction $\hat{\mathbf{b}}$ as $T^{-1}_{1,i}(\hat{\mathbf{b}})=T^{-1}_{1}(\hat{\mathbf{b}}+\theta _i) $, it is not difficult to verify  from  Eq. (\ref{Proj}) that
\begin{eqnarray}
 \sum_i^3 \frac{1}{T}_{1,i}(\hat{\mathbf{b}})
&=&-3\lambda_{2}^{2}\,\int_{0}^{\Lambda}\frac{\textrm{d}\bar{q}}{2\pi}\,\bar{q}\lim_{\omega\to0}\frac{1}{\omega}\left[\textrm{Im}\left(\chi_{xx}^{s}+ \chi_{zz}^{s}\right)\nonumber\right.\\
&&\qquad\quad +\left. \frac{1}{2}\sin^2\varphi\, \textrm{Im}\left(\chi_{xx}^{s}-\chi_{zz}^{s}\right)\right]\,,\label{Proj2}\end{eqnarray}
where $\varphi$ is the  angle that $\hat{\mathbf{b}} $ makes with the normal direction to the SC planes. 
We notice that despite the broken rotational symmetry of the triple-CDW state, the total NMR response is rotationally invariant in  the planes and shows an anisotropic direction along the normal $c$-axis, 
as displayed in Fig. 10.

\section{Transport}

In this section we calculate the optic and thermal conductivities
of the SC phase in the clean limit. The transport calculation for
a $d$-wave order parameter with and without $d-$wave superconductivity
has been done by Yang and Nayak \cite{Yung}. Here, we shall repeat
the calculation for a CDW gap with nodes coexisting with a $s$-wave
SC order parameter. We ignore the effects of scattering centers
like impurities and disorder from the CDW fluctuations motivated by
the two facts: 1) the TMD are very clean materials, and 2) the extremely
low temperatures where the SC phase appears in 2H-TaSe$_{2}$ ($T\lesssim0.1$
K), where conventional thermal disorder in the CDW phase should play
no relevant role in the transport.

The thermal current
is defined by 
$\mathbf{j}^{Q}=\mathbf{j}^{E}-\frac{\mu}{e}\mathbf{j}\,,$\cite{Mahan}
where $\mathbf{j}^{E}$ is the energy current, $\mathbf{j}$ is the
electrical current and $\mu$ is the chemical potential. Experimental
measurements of the thermal conductivity $\kappa$ require zero electric
current flow in the sample, and we may assume that 
$\mathbf{j}^{Q}=\mathbf{j}^{E}$.
The Kubo formulas for frequency dependent 
thermal conductivity, $\kappa(\omega)$, and the optical
conductivity, $\sigma(\omega)$, are \cite{Mahan}: 
\begin{eqnarray}
\kappa_{ij}(\omega)\! & = & \!\!-\frac{1}{\omega
  T}\,\lim_{q\to0}\textrm{Im}\,\Pi_{ij}^{EE}(\mathbf{q},\omega)+TS_{ij}^{2}(\omega)\sigma_{ij}(\omega),\quad 
\label{ThermalConductivity}
\\
\sigma_{ij}(\omega)\! & = &
\!\!-\frac{1}{\omega}\,\lim_{q\to0}\textrm{Im}\,\Pi_{ij}(\mathbf{q},\omega)\,,
\label{sigma}
\end{eqnarray}
where 
\begin{equation}
S_{ij}(\omega)=-\frac{1}{T}\lim_{q\to0}\left[\frac{\textrm{Im}\,\Pi_{ij}^{E}(\mathbf{q},\omega)}{\textrm{Im}\,\Pi_{ij}(\mathbf{q},\omega)}\right]
\label{thermopower}
\end{equation}
is the thermoelectric conductivity (also known as thermopower) 
$S=-\Delta V/\Delta T$,
that measures the current voltage $\Delta V$ produced by a temperature
gradient $\Delta T$, and $\Pi$, $\Pi^{EE}$ and $\Pi^{E}$ are respectively
the electric, thermal and thermoelectric current correlation functions,
which we define in appendix C. The second term in 
Eq. (\ref{ThermalConductivity})
guarantees the zero current flow condition to the charge carriers.

\subsection{Optical conductivity }

To incorporate the magnetic field into Hamiltonian (\ref{H1}), we
proceed with the Peierls substitution 
$\mathbf{k}\to\mathbf{k}-\frac{e}{c}\,\tau_{3}\,\mathbf{A}$.
We assume that the vector potential $\mathbf{A}(\mathbf{k})$ is symmetric
with respect to momentum inversion in the nodal space. For this
reason, we must use the $\tau_{3}$ Pauli matrix, which operates in
the usual Nambu space. Notice that a given
Hamiltonian density for spin $\frac{1}{2}$ fermions in the form 
$\sum_{\sigma}f(\mathbf{k})\psi_{\sigma}^{\dagger}(\mathbf{k})\psi_{\sigma}(\mathbf{k)}$
is equivalently written in the Nambu space as 
\[
\left(\begin{array}{cc}
\psi_{\uparrow}^{\dagger}(\mathbf{k}) & \psi_{\downarrow}(-\mathbf{k})\end{array}\right)\left(\begin{array}{cc}
f(\mathbf{k}) & 0
\\
0 & -f(-\mathbf{k})\end{array}\right)\left(\begin{array}{c}
\psi_{\uparrow}(\mathbf{k})
\\
\psi_{\downarrow}^{\dagger}(-\mathbf{k})\end{array}\right).
\]
The associated matrix above is clearly $\tau_{3}$ if $f$ is a symmetric
function in $k$ and $\tau_{0}$ if $f$ is anti-symmetric. As the
Dirac fermion dispersion is anti-symmetric in the cone, we should
be especially careful with the usual Peierls substitution, since it
introduces an even term ($\propto\tau_{3}\mathbf{A}$), which violates
the odd symmetry of the zero field dispersion dependence with $k$. 
For a given Hamiltonian in the general form:
\[
H=\sum_{\mathbf{k}}\,\epsilon_{\mathbf{k}}\Psi^{\dagger}(\mathbf{k})\,\tau_{0}\eta_{i}\,\Psi(\mathbf{k})\,,
\]
the correct Peierls substitution involves the separation of symmetric
($S$) and anti-symmetric ($A$) components in $\mathbf{k}$, 
\begin{equation}
\epsilon(\mathbf{k})\tau_{0}\to\epsilon^{A}(\mathbf{k}-\frac{e}{c}\tau_{3}\mathbf{A})\,\tau_{0}+\epsilon^{S}(\mathbf{k}-\frac{e}{c}\tau_{3}\mathbf{A})\,\tau_{3}\,,
\label{Peirls}
\end{equation}
where
\begin{eqnarray*}
\epsilon^{S}(\mathbf{k}-\frac{e}{c}\tau_{3}\mathbf{A}) & = &
\frac{1}{2}\left[\epsilon(\mathbf{k}-\frac{e}{c}\tau_{3}\mathbf{A})+\epsilon(-\mathbf{k}-\frac{e}{c}\tau_{3}\mathbf{A})\right],
\\
\epsilon^{A}(\mathbf{k}-\frac{e}{c}\tau_{3}\mathbf{A}) & = &
\frac{1}{2}\left[\epsilon(\mathbf{k}-\frac{e}{c}\tau_{3}\mathbf{A})-\epsilon(-\mathbf{k}-\frac{e}{c}\tau_{3}\mathbf{A})\right].
\end{eqnarray*}
 
Applying this procedure to the Hamiltonian (\ref{H1}), it is easy to
see that the magnetic part of the Hamiltonian is
\[
H_{B}=-\Psi^{\dagger}(\mathbf{k})\left[v_{F}\frac{e}{c}A_{\perp}\tau_{0}\eta_{3}+v_{\Delta}\frac{e}{c}A_{\parallel}\tau_{0}\eta_{1}\right]\Psi(\mathbf{k})\,,
\]
written in terms of $\tau_{0}$ instead \emph{}of \emph{}$\tau_{3}$,
as one could naively expect from the straight substitution 
$\mathbf{k}\to\mathbf{k}-\frac{e}{c}\tau_{3}\,\mathbf{A}$. 

The current density operator $\mathbf{j}(\mathbf{k})=-c\,\nabla_{\mathbf{A}}H$
is given by 
\begin{equation}
\mathbf{j}(\mathbf{k})=\Psi^{\dagger}(\mathbf{k})\left[v_{F}e\,\tau_{0}\eta_{3}\hat{\mathbf{e}}_{\perp}+v_{\Delta}e\,\tau_{0}\eta_{1}\hat{\mathbf{e}}_{\parallel}\right]\Psi(\mathbf{k})\,.
\label{Ecurrent}
\end{equation}
The current-current density correlation function defined in appendix
C is given by:
\begin{eqnarray*}
\Pi_{\perp}(\mathbf{q},i\omega) & = &
\frac{v_{F}^{2}e^{2}}{\beta}\textrm{Tr}\sum_{\mathbf{k},\omega_{n}}\stackrel{\leftrightarrow}{G}_{+}\tau_{0}\eta_{3}\,\stackrel{\leftrightarrow}{G}_{-}\tau_{0}\eta_{3},
\\
\Pi_{\parallel}(\mathbf{q},i\omega) & = &
\frac{v_{\Delta}^{2}e^{2}}{\beta}\textrm{Tr}\sum_{\mathbf{k},\omega_{n}}\stackrel{\leftrightarrow}{G}_{+}\tau_{0}\eta_{1}\,\stackrel{\leftrightarrow}{G}_{-}\tau_{0}\eta_{1},
\end{eqnarray*}
where $\perp$ and $\parallel$ are the normal and parallel directions
to the Fermi surface for a given node (see Fig. 2). Applying the Kubo
formula (\ref{sigma}) to the imaginary part of the correlation functions
above, we find that the optical conductivity is separated into two parts:
the Drude term, 
\begin{eqnarray}
\sigma_{\perp}^{DC}(\omega) & = &
-\frac{v_{F}e^{2}}{2v_{\Delta}}\delta(\omega)\sum_{\sigma^{\prime}}\int_{0}^{\alpha}\textrm{d}\epsilon\,\epsilon\,\left(1-\frac{\Delta_{s}^{2}}{E_{\sigma^{\prime}\mu}^{2}}\right)\nonumber 
\\
 &  & \qquad\qquad\qquad\qquad\times\frac{\partial
   n(E_{\sigma^{\prime}\mu})}{\partial E_{\sigma^{\prime}\mu}},\label{Drudesigma}
\end{eqnarray}
and an extra term due to the interband excitations of the Dirac fermions,
\begin{eqnarray}
\sigma_{\perp}^{AC}(\omega) & = &
\frac{2v_{F}e^{2}}{v_{\Delta}}\,\frac{\Delta_{s}^{2}}{\omega^{2}}\left[n\left(-\frac{|\omega|}{2}\right)-n\left(\frac{|\omega|}{2}\right)\right]\nonumber 
\\
 &  &
 \quad\times\theta\left(|\omega|-2\sqrt{\mu^{2}+\Delta_{s}^{2}}\right)\nonumber 
\\
\nonumber 
\\ &  &
+\frac{v_{F}e^{2}}{2v_{\Delta}}\hbar\omega\nu_{0}\left(1-\frac{4\mu^{2}}{\omega^{2}}\right)\nonumber 
\\
 &  & \times\left\{ \theta\left(|\mu|
     -\frac{|\omega|}{2}\right)\frac{1}{\Theta_{-}}\left[n(E_{0,|\mu|})-n(E_{0,-|\mu|})\right]\right.\nonumber 
\\
 &  &
 \quad-\theta\left(\frac{|\omega|}{2}-\sqrt{\mu^{2}+\Delta_{s}^{2}}\right)\nonumber 
\\
 &  &
 \quad\qquad\left.\times\frac{1}{\Theta_{+}}\left[n(E_{0,\mu})-n(-E_{0,-\mu})\right]\right\} ,
\label{sigmaMuPerp}
\end{eqnarray}
\begin{eqnarray*}
\quad & \stackrel{T\to0}{\longrightarrow} &
\frac{v_{F}e^{2}}{2v_{\Delta}}\left[\left(1-\frac{4\mu^{2}}{\omega^{2}}\right)\frac{\omega\nu_{0}}{\Theta_{+}}+\frac{4\Delta_{0}^{2}}{\omega^{2}}\right]\qquad
\\
 &  & \qquad\times\theta\left(|\omega|-2\sqrt{\mu^{2}+\Delta_{s}^{2}}\right)
\end{eqnarray*}
where 
\begin{equation}
\nu_{0}\equiv\frac{\omega}{2}\,\sqrt{1-\frac{4\Delta_{s}^{2}}{\omega^{2}-4\mu^{2}}}\,,
\label{nu_0}
\end{equation}
and 
\[
\Theta_{\pm}=\left|(|\nu_{0}|-\mu)E_{0,\mu}\pm(|\nu_{0}|+\mu)E_{0,-\mu}\right|\,,
\]
with $E_{0,\pm\mu}^{2}=(|\nu_{0}|\pm\mu)^{2}+\Delta_{s}^{2}$. In order to
calculate the the parallel component $\sigma_{\parallel}$ we just have to
replace $v_{F}$ by $v_{\Delta}$. For $\mu=0$, the
interband conductivity is given by:
\begin{equation}
\sigma_{\perp}^{AC}(\omega)=\frac{e^{2}v_{F}}{2v_{\Delta}}\left(1+\frac{4\Delta_{s}^{2}}{\omega^{2}}\right)\left|1-2n\,\left(\frac{\omega}{2}\right)\right|\theta(|\omega|-2\Delta_{s})\,.
\label{sigmaAC}
\end{equation}

The conductivity (\ref{sigmaMuPerp}) is considerably simpler in the
normal CDW phase. Setting the gap $\Delta_{s}$ to zero, we have 
$\nu_{0}\to\omega/2$
and $E_{0,\pm\mu}\to\left|\frac{|\omega|}{2}\pm\mu\right|$, leading to:
\begin{eqnarray}
\sigma_{\perp CDW}^{AC}(\omega)\!\! & = &
\frac{v_{F}e^{2}}{2v_{\Delta}}\left[n\left(-\frac{|\omega|}{2}+\mu\right)-n\left(\frac{|\omega|}{2}+\mu\right)\right]\nonumber 
\\
& \stackrel{T\to0}{\longrightarrow} &
\frac{v_{F}e^{2}}{2v_{\Delta}}\,\theta\!\left(\frac{|\omega|}{2}-|\mu|\right)\,.
\label{sigmaMuCDW}
\end{eqnarray}
Analogously, the Drude part of the conductivity becomes 
\begin{eqnarray}
\sigma_{\perp CDW}^{DC}(\omega)\!\! & = &
\frac{v_{F}e^{2}\beta}{8v_{\Delta}}\delta(\omega)\sum_{\sigma^{\prime}}\int_{0}^{\alpha}\textrm{d}\epsilon\,\epsilon\,\nonumber 
\\
&  &
\qquad\qquad\times\textrm{sech}^{2}\left(\beta\frac{\epsilon+\sigma^{\prime}\mu}{2}\right)
\nonumber 
\\
 & \stackrel{T\to0}{\longrightarrow} &
 \frac{v_{F}e^{2}}{2v_{\Delta}}\,\delta(\omega)\times\left\{ 
\begin{array}{cl}
\frac{2 \ln(2)}{\beta} & \,,\,\textrm{for}\:\mu=0
\\
\\|\mu| & \,,\,\textrm{for}\:\mu\neq0\,.\end{array}\right.\quad
\label{sigmaDCMu}
\end{eqnarray}

Notice that in the absence of SC we find that $\sigma^{DC}(T \to 0)$ 
is constant and proportional to $\mu$. 
In the SC case, Eq. (\ref{Drudesigma}) shows that $\sigma^{DC}(T \to 0)$
vanishes independently of the pocket size (as shown in Fig. 11). 
The presence of a Drude conductivity, $\sigma^{DC} \propto \delta(\omega)$,  
results from an infinite electron mean free path 
due to the absence of scattering
centers. If we consider that the electrons in TaSe$_2$ have a finite 
scattering rate, $\Gamma=1/\tau(\omega)$ \cite{dordevic3}, the Drude 
peak will be broadened around $\omega=0$. 
The normal transport in the presence of an order parameter with nodes 
(like the CDW, as in  our case) in the dirty limit, is given in Ref.~[\onlinecite{Yung}].

\begin{figure}[h]
\begin{center}$\!\!\!\!\!\!\!\!$\includegraphics[%
  scale=0.81]{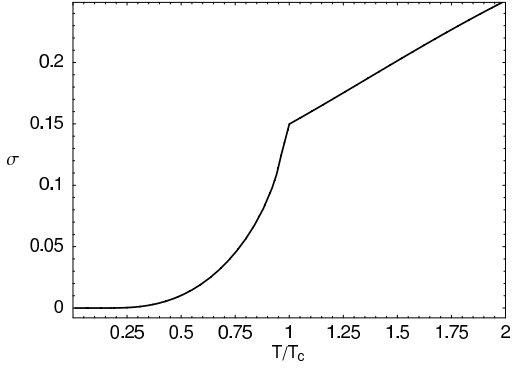}$\qquad$\end{center}
\caption{{\small Temperature dependence of the Drude conductivity integrated in $\omega$, for $g/g_{c}=1.1$
and $\mu/\alpha=0.1$. $\sigma$ in units of $v_{F}e^{2}\alpha/(2v_{\Delta})$. }}
\end{figure}

Photon absorption involves quasiparticle excitations and results
in the formation of in-phase currents with the electric field \cite{Parks}.
The absorption rate is therefore proportional to the real part of
the conductivity. In conventional superconductors there is no absorption
at $T=0$ in the infrared region where the photons with energy $\omega<2\Delta_{s}$
cannot break a Cooper pair. At finite temperature, the excitation channels
are gradually recovered and photons with energy smaller than $2\Delta_{s}$
have a finite probability of being absorbed. We should stress that
the coherence factors in those superconductors (say, BCS type) are
finite only in the dirty case, where the processes conserve energy
but do not conserve momentum. The first important distinction between
traditional BCS superconductors to the ones discussed here is the presence
of \emph{two} bands, resembling the spectrum of small gap semiconductors
(see Fig. 3). In the nodal liquid superconductor, made out of Cooper pairs of 
Dirac fermions, the absorption process comprehends the excitation of an electron
from the lower to the upper band, transferring energy equal to the
the photon energy $\omega$ but with no momentum transfer. In the
situation where the lower band is completely filled ($\mu=0$), there
are no thermal channels of quasiparticle excitations (since the thermally
excited electrons occupy the upper band, where there is no absorption
due to momentum conservation) and the photon is absorbed only
when its energy is sufficient to break a pair ($\omega>2\Delta_{s}$),
producing quasiparticle excitations directly from the condensate 
(pair breaking channels).
When the system exhibits particle-hole symmetry, the absorption
is independent of the temperature in the infrared for $\omega<2\Delta_{s}$.

\begin{figure}[]
\begin{center}$\!\!\!\!\!\!\!\!\!\!\!\!\!\!\!\!\!\!\!\!$\includegraphics[%
  scale=0.81]{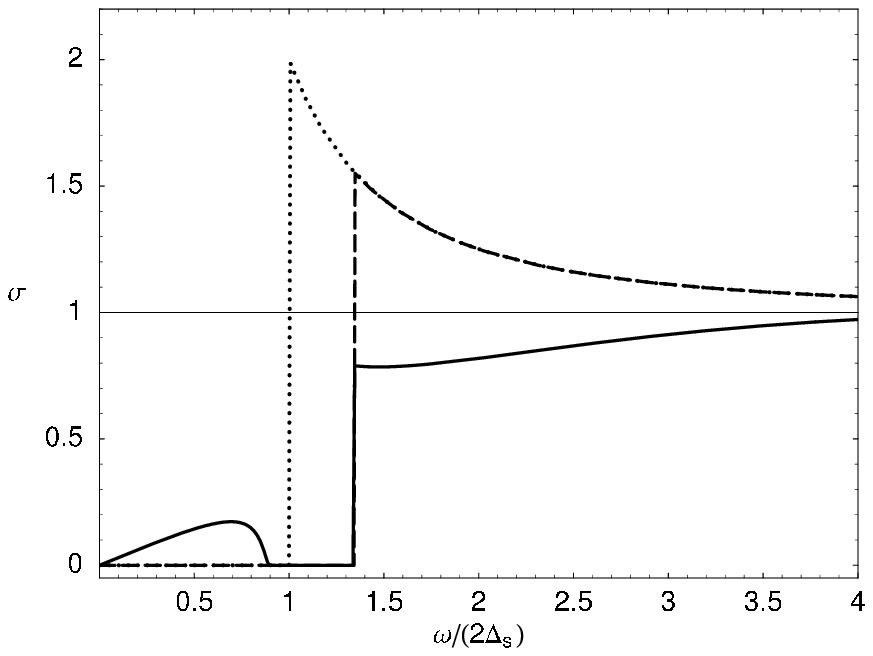}$\,$\\
$\!\!\!\!\!\!\!\!\!\!\!\!\!\!\!\!\!\!\!\!$\includegraphics[%
  scale=0.81]{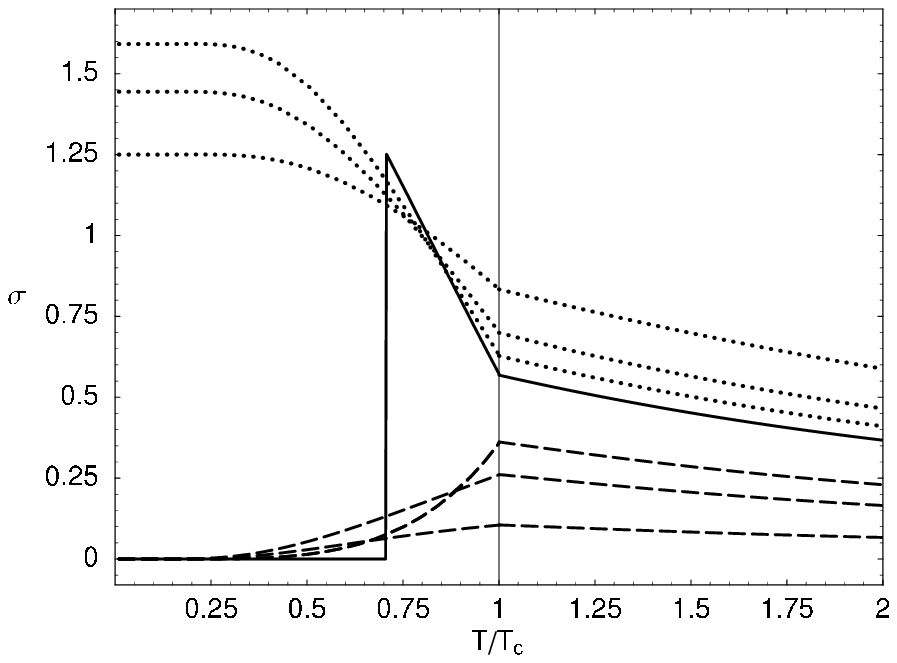}$\quad\,$\end{center}

\caption{{\small Top: optical conductivity} \emph{\small $\sigma_{\perp}$}
{\small vs. frequency. For $|\mu|/(2\Delta_{s})=0.9$: dashed line ($T=0$)
and solid ($k_{B}T/(2\Delta_{s})=1.2$); dotted line: $\mu=0$ and
$T=0$. Bottom: optical conductivity $\sigma_{\perp}$ vs. temperature,
for $g/g_{c}=1.1$ and $|\mu|/\alpha=0.1$. Dashed lines: $0.4\Delta_{0\mu}<\omega<1.4\Delta_{0\mu}$;
solid: $\omega=2.3\Delta_{0\mu}$; dotted: $2.8\Delta_{0\mu}<\omega<4\Delta_{0\mu}$.
In both plots, $\sigma$ is in units of $v_{F}e^{2}/(2v_{\Delta})$}}
\end{figure}

The second important distinction between the TMD and BCS superconductors, 
is that the optical conductivity shows
an anomalous absorption edge in $\omega=2\sqrt{\mu^{2}+\Delta_{s}^{2}}$
(see Fig. 12). This energy corresponds to the optical gap of the two
bands shown in Fig. 3. The presence of the edge is a consequence of
the broken lattice inversion symmetry in the CDW phase,
which affects the coherence factors of the infrared conductivity.
When the particle-hole
symmetry is lost by shifting the chemical potential from the vertex
of the Dirac cone, new thermal channels of quasiparticle excitations
emerge, giving rise to an absorption peak in the infrared. To see
this effect, we illustrate in  Fig. 13 the thermal excitation process of the
hole-like branch, where the photons with energy smaller than $2|\mu|$
are able to promote the thermally excited electrons occupying the
empty states on the top of the lower band to the upper band. 
As in the case of superfluid He$^3$, the superconductor is an electronic liquid composed
of two {}``fluids'', where there is a one-to-one correspondence
between the excited states in the SC and in the normal phases. The
thermal fluctuations promote electrons from the condensate to the empty
states above the pocket Fermi surface of the hole-like branch. The optical channels
of absorption through the thermally excited electrons are therefore
limited to the window $|\omega|\leq2|\mu|$ (in the clean limit), as
shown in Fig. 12 (top) and Fig. 13. 

\begin{figure}
\begin{center}\vspace{0.2cm}\includegraphics[%
  scale=0.39]{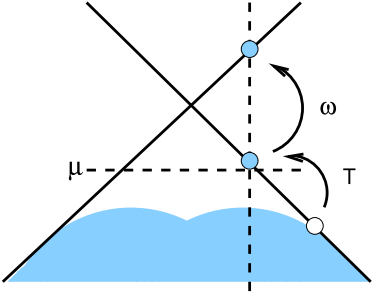}\end{center}

\caption{{\small Schematic representation of the photon absorption process
in the channel of thermal excitations of the condensate, within the
absorption window $|\omega|<2|\mu| $ of the hole-like branch (see Fig. 3). $\omega$ is the photon frequency,
 $T$ represents the thermal excitations and $\mu$ indicates the Fermi level.}}
\end{figure}

The temperature dependence of the optical conductivity, displayed in
Fig. 12 (bottom), shows a clear distinction between the two absorption
channels. The dashed lines represent the thermal channels, which vanish
at $T=0$. The dotted lines indicate the pair breaking channels. 
These channels depend on the number
of electrons in the condensate and are more effective as the temperature
is reduced. The solid line in the same figure represents a pair breaking
channel which is abruptly suppressed by lowering the temperature.
This is understood by noting that the optical gap $2\sqrt{\mu^{2}+\Delta_{s}^{2}(T)}$
{[}see Fig. 12 (top){]} displaces the absorption edge towards
the ultraviolet as the temperature is reduced. In this situation,
we expect that some of the absorption channels, at a given energy slightly
to the right of the edge, will be abruptly suppressed if the temperature
is sufficiently reduced, \emph{i.e}. if the edge is sufficiently displaced
to the right in Fig. 12.

\subsubsection{Spectral weight}

According to the $f-$sum rule one should have
\begin{eqnarray}
\int_{0}^{\infty}\sigma(\omega)\,\textrm{d}\omega & = & \frac{\pi
  ne^{2}}{2m}\,,
\label{sumR}
\end{eqnarray}
and therefore, the area {}``under'' the curves $\sigma^{DC}+\sigma^{AC}$ is conserved
in the normal and in the SC phases. In the SC phase, however, there
is a {}``missing'' area in comparison to the normal phase. The difference
between the two areas corresponds to the $\omega=0$ spectral weight,
responsible for the diamagnetic supercurrents in the Meissner effect
\cite{tinkham}. This part of the spectral weight (which properly
defines a superconductor) depends on a different order of limits between
$\omega$ and $q$, and does not appear explicitly in the calculation.
Thus, a required condition for superconductivity is 
\[
\int_{0}^{\infty}[\sigma_{s}^{DC}(\omega)+\sigma_{s}^{AC}(\omega)]\,\textrm{d}\omega<\int_{0}^{\infty}[\sigma_{n}^{DC}(\omega)+\sigma_{n}^{AC}(\omega)]\,\textrm{d}\omega\,.
\] From now on, we call the difference between the $n$ and $s$
areas as the \emph{Meissner spectral weight}. 

It is not difficult to see that for $\mu=0$ 
at zero temperature we have $\sigma_{s}^{DC}=\sigma_{n}^{DC}=0$,
and that the curves in the AC sector have \emph{exactly} the same
area. This behavior is depicted in Fig. 14 for different values of $\mu$,
showing an anomalous suppression of the Meissner spectral weight at
low temperatures for small $\mu$. A superficial analysis would indicate
that there is no spectral weight due to the condensate and therefore
the superconductivity is not stable. This analysis, however, is incompatible
with the thermodynamic verification that there is a finite zero temperature
critical field $H_{c}(0)$ (see Fig. 7), resulting in a finite condensation
energy. 
The origin of the problem has connections with the spectral weight
shift from the high to the low energy states of the band as the temperature
is reduced, which has been observed experimentally in TaSe$_{2}$
\cite{vescoli}. 
In this compound, part of the spectral weight around
60 meV ($\sim$ of the order of the cone cut-off) at $300$ K is displaced
towards the infrared at temperatures of the order of the SC phase transition.
Apparently, the opening of the gap ``attracts states'' beyond the cone
approximation. In the lowest order, the non-linear states
in the CDW spectrum yield $\epsilon_{\mathbf{k}}\propto(k-\frac{e}{c}A)^{2}$.
These states are the only ones that contribute to the diamagnetism,
which results from terms $\propto A^{2}$ in the energy. We conclude
that the cone approximation excludes the {}``diamagnetic'' states
of the band, and for this reason the $f$-sum rule is not able to
correctly incorporate the diamagnetic spectral weight, specially at
low temperature, where the contribution of the high energy states
is more pronounced. The zero field properties which are not directly
related to the Meissner effect, however, are not so sensitive to the
absence of the high energy states and give satisfactory results within
the cone approximation. This analysis is confirmed later in sec.
VII, when we discuss the Meissner effect in the London limit.

\begin{figure}[b]
\begin{center}$\!\!\!\!\!\!\!\!\!\!\!$\includegraphics[%
  scale=0.81]{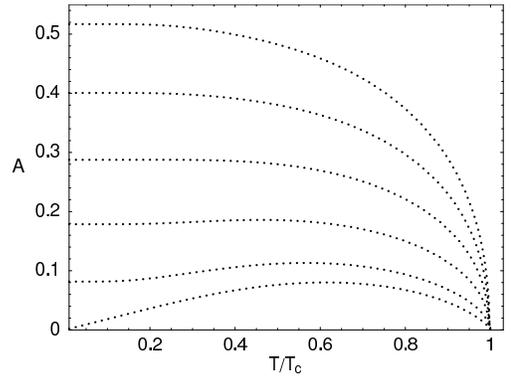}$\qquad\:$\vspace{0.1cm}\end{center}

\caption{{\small Meissner spectral weight $A$ as a function of temperature.
Curves drawn for $0\leq |\mu| /\alpha\leq0.15$, from the bottom to the
top, in fixed intervals of $0.03$. $A$ in units of $v_{F}e^{2}\alpha/(2v_{\Delta})$,
with $g/g_{c}=1.1$.}}
\end{figure}

\subsection{Thermal conductivity}

The energy current is a conserved quantity defined by the non-diagonal
components of the momentum-energy tensor $T_{\,\,0}^{i}$, defined
as \cite{Peskin}
\begin{equation}
T_{\,\,\nu}^{\mu}\equiv\frac{\partial\mathcal{L}}{\partial(\partial_{\mu}\Psi)}\partial_{\nu}\Psi-\mathcal{L}\delta_{\,\,\nu}^{\mu}\,.
\label{TensorEM}
\end{equation}
According to the usual relation 
$H=\frac{\partial\mathcal{L}}{\partial(\partial_{0}\Psi)}\partial_{0}\Psi-\mathcal{L}$,
the Lagrangian associated to the Hamiltonian (\ref{H1}) in the real
space representation is
\begin{eqnarray}
\mathcal{L} & = &
\Psi^{\dagger}(x)\left[ic\tau_{3}\eta_{0}\partial_{0}-i\hbar
  v_{F}\tau_{0}\eta_{3}\partial_{3}\right.\nonumber 
\\
 &  &
 \left.-iv_{\Delta}\tau_{0}\eta_{1}\partial_{1}-\Delta_{s}\tau_{1}\eta_{2}+\mu \tau_3\eta_0\right]\Psi(x)\,,
\end{eqnarray}
where $c\partial_{0}\equiv i\partial_{\tau}$ with $\tau$ as the
imaginary time. The conserved energy current $\mathbf{j}^{E}(x)=cT_{\,\,0}^{i}$
gives
\begin{eqnarray*}
\mathbf{j}^{E}(x) & = & \frac{\partial\mathcal{L}}{\partial(\partial_{i}\Psi)}c\partial_{0}\Psi\\
 & = &
 \Psi^{\dagger}(x)\left[v_{F}\tau_{0}\eta_{3}\hat{\mathbf{e}}_{\perp}+v_{\Delta}\tau_{0}\eta_{1}\hat{\mathbf{e}}_{\parallel}\right]\partial_{\tau}\Psi(x)\,,
\end{eqnarray*}
or equivalently
\begin{eqnarray}
\mathbf{j}^{E}(\mathbf{q},\tau) & = &
-\sum_{\mathbf{k}}\Psi^{\dagger}(\mathbf{k}-\mathbf{q}/2,\tau)\left[v_{F}\tau_{0}\eta_{3}\hat{\mathbf{e}}_{\perp}\right.\nonumber 
\\
 &  &
 \quad\left.+v_{\Delta}\tau_{0}\eta_{1}\hat{\mathbf{e}}_{\parallel}\right]\stackrel{\leftrightarrow}{\omega}_{\mathbf{k}+\mathbf{q}/2}\Psi(\mathbf{k}+\mathbf{q}/2,\tau)\,,
\qquad\label{thermalJ}
\end{eqnarray}
where the time-evolution of the Dirac fermions $\Psi$ is
\[
\Psi(\mathbf{q},\tau)=\textrm{e}^{-\tau\stackrel{\leftrightarrow}{\omega}_{\mathbf{q}}}\Psi(\mathbf{q)}\,,\]
with $\stackrel{\leftrightarrow}{\omega}$ defined in Eq. (\ref{H1}). 

We are interested in the diagonal components of the current-current polarizations
$\Pi_{11}^{EE}\equiv\Pi_{\perp}^{EE}$ and $\Pi_{11}^{E}\equiv\Pi_{\perp}^{E}$
given by:
\begin{eqnarray*}
\Pi_{\perp}^{EE}(\mathbf{q},i\omega) & = &
\frac{v_{F}^{2}}{\beta}\textrm{Tr}\sum_{\mathbf{k},\omega_{n}}\stackrel{\leftrightarrow}{G}_{+}\tau_{0}\eta_{3}\stackrel{\leftrightarrow}{\omega}_{+}\,\stackrel{\leftrightarrow}{G}_{-}\tau_{0}\eta_{3}\stackrel{\leftrightarrow}{\omega}_{-},
\\
\Pi_{\perp}^{E}(\mathbf{q},i\omega) & = &
\frac{v_{F}^{2}e}{\beta}\textrm{Tr}\sum_{\mathbf{k},\omega_{n}}\,\stackrel{\leftrightarrow}{G}_{+}\tau_{0}\eta_{3}\stackrel{\leftrightarrow}{\omega}_{+}\,\stackrel{\leftrightarrow}{G}_{-}\tau_{0}\eta_{3}\,.
\end{eqnarray*}
We find that 
\begin{eqnarray}
\textrm{Im}\,\Pi_{\perp}^{EE}(0,\omega)\!\! & = &
\!\!\frac{v_{F}\omega}{2v_{\Delta}}\,\delta(\omega)\sum_{\sigma^{\prime}=\pm1}\int_{0}^{\alpha}\textrm{d}\epsilon\,\epsilon
E_{\sigma^{\prime}\mu}^{2}\nonumber 
\\
 &  &
 \qquad\qquad\times\left(1-\frac{\Delta_{s}^{2}}{E_{\sigma^{\prime}\mu}^{2}}\right)\frac{\partial n(E_{\sigma^{\prime}\mu})}{\partial E_{\sigma^{\prime}\mu}}\nonumber 
\\
 &  &
 \!\!\!\!-\frac{v_{F}}{2v_{\Delta}}\omega^{2}\nu_{0}\left(1-\frac{4\mu^{2}}{\omega^{2}}\right)E_{0,\mu}E_{0,-\mu}\nonumber 
\\
 &  & \!\!\!\!\times\!\left\{
   \theta\!\left(|\mu|-\frac{|\omega|}{2}\right)\!\frac{1}{\Theta_{-}}\!\left[n(E_{0,|\mu|})-n(E_{0,-|\mu|})\right]\right.\nonumber 
\\
 &  &
 \quad+\theta\!\left(\frac{|\omega|}{2}-\sqrt{\mu^{2}+\Delta_{s}^{2}}\right)\nonumber 
\\
 &  &
 \qquad\quad\left.\times\frac{1}{\Theta_{+}}\left[n(E_{0,\mu})-n(-E_{0,-\mu})\right]\right\} \nonumber 
\\
 &  &
 \!\!\!\!+\frac{v_{F}\omega}{2v_{\Delta}}\,\Delta_{s}^{2}\left[n\left(-\frac{|\omega|}{2}\right)-n\left(\frac{|\omega|}{2}\right)\right]\nonumber 
\\
 &  & \qquad\qquad\times\theta\left(|\omega|-2\sqrt{\mu^{2}+\Delta_{s}^{2}}\right)\,,\label{PiEEMuText}\end{eqnarray}
and 
\begin{eqnarray}
\textrm{Im}\,\Pi_{\perp}^{E}(0,\omega) & = &
\!\!\frac{v_{F}e}{4v_{\Delta}}\omega^{2}|\nu_{0}|\left(1-\frac{4\mu^{2}}{\omega^{2}}\right)\nonumber 
\\
 &  & \times\left\{
   \theta\left(|\mu|-\frac{|\omega|}{2}\right)\left[E_{0,\mu}-E_{0,-\mu}\right]\right.\nonumber 
\\
 &  & \qquad\qquad\times\frac{1}{\Theta_{-}}\left[n(E_{0,\mu})-n(E_{0,-\mu})\right]\nonumber \\
 &  &
 \!\!-\theta\left(\frac{|\omega|}{2}-\sqrt{\mu^{2}+\Delta_{s}^{2}}\right)\left[E_{0,\mu}+E_{0,-\mu}\right]\nonumber 
\\
 &  &
 \qquad\quad\left.\times\frac{1}{\Theta_{+}}\left[n(E_{0,\mu})-n(-E_{0,-\mu})\right]\right\} \nonumber 
\\
 &  &
 \!\!+\frac{v_{F}e}{v_{\Delta}}\Delta_{s}^{2}\left[n\left(-\frac{|\omega|}{2}\right)-n\left(\frac{|\omega|}{2}\right)\right]\nonumber 
\\
 &  &
 \qquad\qquad\times\theta\left(|\omega|-2\sqrt{\mu^{2}+\Delta_{s}^{2}}\right),
\label{ThermopowerMu}
\end{eqnarray}
where $ \nu_0$ and $ E_{0,\sigma \mu}$ are defined as in Eq. (\ref{sigmaMuPerp}). 
In contrast to the thermal polarization, the thermoelectric one does
not have a Drude part. The thermal conductivity follows from a straightforward
substitution of the previous results (\ref{sigmaMuPerp}), (\ref{PiEEMuText})
and (\ref{ThermopowerMu}) into the Kubo formula (\ref{ThermalConductivity}). 

Let us analyze these results for $\mu=0$. We have
\begin{eqnarray}
\frac{1}{\omega}\textrm{Im}\,\Pi_{\perp}^{EE}(0,\omega) & = &
\frac{v_{F}}{v_{\Delta}}\,\delta(\omega)\int_{\Delta_s}^{E_\Lambda}\textrm{d}
E^{3}\left(1-\frac{\Delta_{s}^{2}}{E^{2}}\right)\nonumber 
\\
 &  & \qquad\qquad\quad\times\frac{\partial n(E)}{\partial E}\nonumber 
\\
 &  &
 +\frac{v_{F}}{2v_{\Delta}}\left(\frac{\omega}{2}\right)^{2}\left(1+\frac{4\Delta_{s}^{2}}{\omega^{2}}\right)\nonumber 
\\
 &  & \quad\times\left[1-2n\left(\frac{|\omega|}{2}\right)\right]\theta(\omega-2\Delta_{s})\,,\nonumber \\
\label{PiEEAC}
\end{eqnarray}
and 
\begin{eqnarray}
\frac{1}{\omega}\textrm{Im}\,\Pi_{\perp}^{E}(0,\omega) & = &
\frac{ev_{F}}{2v_{\Delta}}\,\frac{\omega}{2}\left(1+\frac{4\Delta_{s}^{2}}{\omega^{2}}\right)
\label{PiECAC}
\\
 &  &
 \quad\times\left[1-2n\!\left(\frac{|\omega|}{2}\right)\right]\theta(|\omega|-2\Delta_{s})\,.
\nonumber \end{eqnarray}
 Replacing Eq. (\ref{sigmaAC}) and (\ref{PiECAC}) into Eq. (\ref{thermopower}),
the $\mu=0$ thermopower yields 
\begin{equation}
S_{\perp}=-\frac{1}{T}\,\frac{\textrm{Im}\,\Pi_{\perp}^{E}(0,\omega)}{\textrm{Im}\,\Pi_{\perp}(0,\omega)}=
\frac{\omega}{2eT}\,.\label{thermoP}
\end{equation}
 
Substituting Eq. (\ref{sigmaAC}), (\ref{PiEEAC}) and (\ref{thermoP})
into (\ref{ThermalConductivity}), we find that the only contribution
comes from the Drude term 
\begin{equation}
\kappa_\perp(\omega)=-\frac{v_{F}}{v_{\Delta}T}\,\delta(\omega)\int_{\Delta_{s}}^{E_\Lambda}\textrm{d}E\,
E^{3}\left(1-\frac{\Delta_{s}^{2}}{E^{2}}\right)\frac{\partial n(E)}{\partial
  E}\,,
\end{equation}
where $\kappa^{AC}=0$ for zero $\mu$. When the system exhibits particle-hole symmetry,
the exact cancellation of the interband contributions to the thermal
conductivity is due to the fact that the total heat carried by a
particle-hole pair is zero. The argument is the following \cite{Yung}:
the interband excitation process involves the annihilation of an electron
with negative energy in the lower band, and the creation of a particle
with positive energy $-E_{\mathbf{k}}+ \omega=+E_{\mathbf{k}}$
in the upper band, where $\omega$ is the photon energy and $-E_{\mathbf{k}}$
is the energy of the annihilated electron. Destroying a particle with
negative energy, momentum $\mathbf{k}$ and charge $e$ is equivalent
to create a hole with momentum $-\mathbf{k}$ and charge $-e$ at
the energy cost $+E_{\mathbf{k}}$. The energy current carried
by the quasiparticle formed by the particle-hole pair is 
$\mathbf{k}\, E_{\mathbf{k}}+(-\mathbf{k})(E_{\mathbf{k}})=0$.
On the other hand, the charge current is finite, $\mathbf{k}\, e+(-\mathbf{k})(-e)=2e\mathbf{k}$,
explaining why the quasiparticles are able to transport charge but
not heat when the pocket is absent. 

When the particle-hole symmetry symmetry is lost, the thermal current
due to the pair breaking channels is equal to $E_{\mathbf{k},-\mu}(\mathbf{k})+(E_{\mathbf{k},\mu})(-\mathbf{k})$,
or equivalently to $-2\mu\mathbf{k}$ in the normal CDW phase, when
the ground state electrons are promoted to the upper band. As a second
effect, the thermal channels of quasiparticle production give rise
to an infrared peak for $|\omega|<2|\mu|$ as shown in Fig. 15 (top),
analogously to the optical conductivity. In contrast with the charge
transport, however, the amount of heat carried by the quasiparticles
is of the order of the pocket energy and vanishes at $\mu=0$. The
temperature dependence of $\kappa$ is shown in Fig. 15 (bottom).
The solid lines represent the thermal channels of quasiparticle excitation,
while the dotted lines indicate the pair breaking channels. As in case of
the optical conductivity, some of the latter channels which are slightly
above the optical gap energy $\omega_{0}=2\sqrt{\mu^{2}+\Delta_{s}^{2}}$
are suppressed at low temperatures (see Fig. 15 ). At $T=0$ the thermal
conductivity is zero for $|\omega|<$ $\omega_{0}$, and infinity
for $|\omega|>$$\omega_{0}$.

\begin{figure}
\begin{center}$\!\!\!\!\!\!\!\!\!\!\!\!\!\!\!\!\!\!$\includegraphics[%
  scale=0.81]{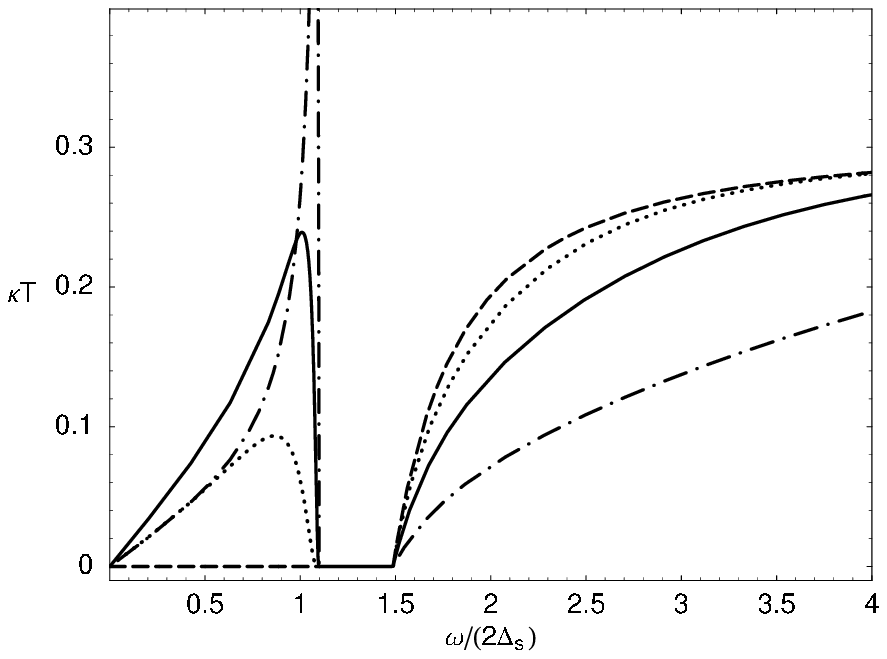} \\
$\!\!\!\!\!\!\!\!\!\!\!\!\!\!\!\!\!\!$\includegraphics[%
  scale=0.81]{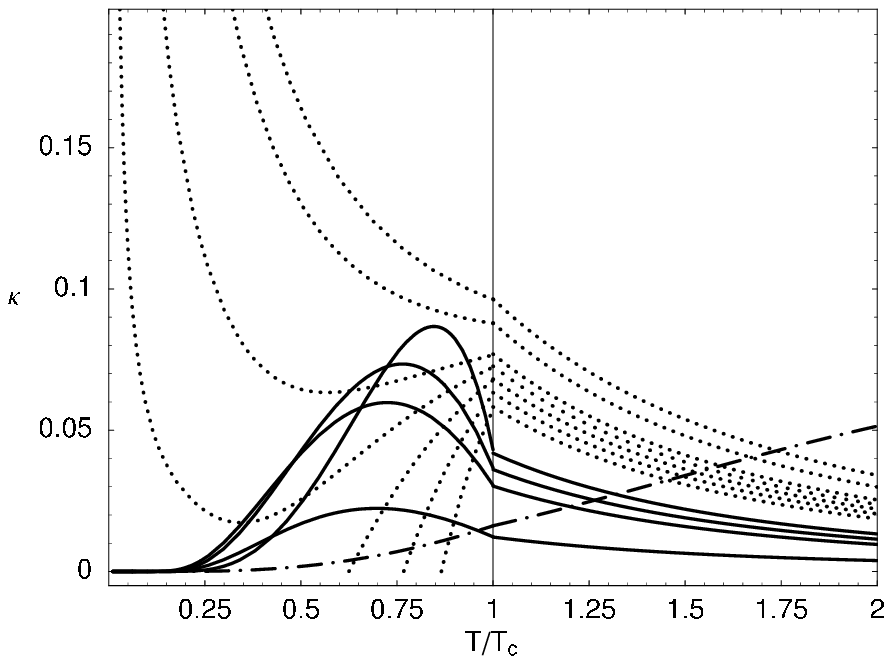}\end{center}

\caption{{\small Top: thermal conductivity $\times T$ vs. frequency. $\kappa T$
is in units of $(2v_{F}/v_{\Delta})\Delta_{s}^2$ and $\omega$
in units of $2\Delta_{s}$, with $\mu=2.2\Delta_{s}$. Dashed line:
$T\to0$ limit; dotted ($k_{B}T=\frac{1}{2}\Delta_{s}$); solid: ($k_{B}T=\Delta_{s}$);
dot-dashed: ($k_{B}T=\frac{5}{2}\Delta_{s}$). Bellow: thermal conductivity
dependency with temperature. We have set $\kappa$ in units of $v_{F}k_{B}\alpha/(2v_{\Delta})$,
$g/g_{c}=1.1$ and $|\mu/\alpha=0.1|$. Solid lines: $0.4\Delta_{0\mu}<\omega<1.4\Delta_{0\mu}$;
dotted: $2\Delta_{0\mu}<\omega<4\Delta_{0\mu}$.} {\small The dot-dashed
line is the Drude thermal conductivity integrated in  $\omega$, with units of $\frac{1}{5}v_{F}k_{B}\alpha^2/(2v_{\Delta})$.}}
\end{figure}

Let us verify the normal CDW properties ($\Delta_{s}=0$) in the transport.
The thermoelectric spectral function (\ref{ThermopowerMu}) is given
by 
\begin{eqnarray*}
\frac{1}{\omega}\textrm{Im}\,\Pi_{\perp CDW}^{E}(0,\omega) & = &
\frac{v_{F}e}{2v_{\Delta}}\frac{\omega}{2}\left[n\!\left(-\frac{|\omega|}{2}+\mu\right)\right.
\\
 &  & \qquad\qquad\left.-n\!\left(\frac{|\omega|}{2}+\mu\right)\right]\,.
\end{eqnarray*}
Comparing the expression above with the optical conductivity of the
normal phase (\ref{sigmaMuCDW}), the thermoelectric coefficient yields
\[
S_{\perp CDW}(\omega)= \frac{\omega}{2eT}\,,
\]
as in the SC particle-hole symmetric case (\ref{thermoP}). Returning
to Eq. (\ref{PiEEMuText}), and setting $\Delta_{s}=0$, we have
\begin{eqnarray*}
\frac{1}{\omega}\textrm{Im}\,\Pi_{\perp CDW}^{EE}(0,\omega)\! & =\!\! &
\frac{v_{F}}{2v_{\Delta}}\left[\left(\frac{\omega}{2}\right)^{2}-\mu^{2}\right]
\\
 &  &
 \times\left[n\!\left(-\frac{|\omega|}{2}+\mu\right)-n\!\left(\frac{|\omega|}{2}+\mu\right)\right]\,.
\end{eqnarray*}
The thermal conductivity is given by: 
\begin{eqnarray*}
\kappa_{CDW}(\omega) & = &
\kappa_{CDW}^{DC}(\omega)+\kappa_{CDW}^{AC}(\omega)\,,
\end{eqnarray*}
where
\begin{eqnarray}
\kappa_{\perp CDW}^{DC}(\omega)\!\! & = &\! -\frac{v_{F}\delta(\omega)}{2v_{\Delta}T}\,\sum_{\sigma^{\prime}=\pm1}\int_{0}^{\alpha}\textrm{d}\epsilon\,\epsilon E_{\sigma^{\prime}\mu}^{2}\nonumber \\
 &  & \qquad\qquad\qquad\quad\times\frac{\partial
 n(E_{\sigma^{\prime}\mu})}{\partial E_{\sigma^{\prime}\mu}}\nonumber 
\\
 & \stackrel{T\to0}{\longrightarrow} & \frac{v_{F}k_{B}}{2v_{\Delta}\beta^{2}}\,\delta(\omega)\times\!\left\{ \begin{array}{cl}
9\zeta(3) & \,,\textrm{for}\,\mu=0
\\
\\\frac{\pi^{2}}{3}\beta|\mu| &
 \,,\textrm{for}\,\mu\neq0\,,\end{array}\right.\quad
\label{kappaDCMu}
\end{eqnarray}
and
\begin{eqnarray}
\kappa_{\perp CDW}^{AC}(\omega)\!\!\! & = &
\!\!\mu^{2}\frac{v_{F}}{2v_{\Delta}T}\left[n\!\left(-\frac{|\omega|}{2}+\mu\right)-n\!\left(\frac{|\omega|}{2}+\mu\right)\right]\nonumber 
\\
 & \stackrel{T\to0}{\longrightarrow} &
 \mu^{2}\frac{v_{F}}{2v_{\Delta}T}\,\theta\!\left(\frac{|\omega|}{2}-|\mu|\right)\,.
\label{kappaACbeta0}
\end{eqnarray}

The verification of the Wiedmann-Franz (WF) law can be done in two
cases. For $\mu=0$, despite the optical conductivity is dominated
in the low temperature region by the interband conductivity, 
\begin{eqnarray}
\sigma_{\perp CDW}(\omega,T\to0) & = & \frac{v_{F}e^{2}}{2v_{\Delta}}\,\tanh\!\left(\frac{\beta|\omega|}{4}\right)\nonumber \\
 &  & +\ln(2)\,\frac{v_{F}e^2}{v_{\Delta}\beta}\,\delta(\omega)\,,
\label{sigmaCDWT=0}
\end{eqnarray}
the $|\omega|/(k_{B}T)\ll1$ limit is dominated by the Drude part.
Comparing the expression above with Eq. (\ref{kappaDCMu}) for $\mu=0$,
we see that the CDW phase obeys the temperature dependence of the
WF law
\begin{equation}
\lim_{T\to0}\frac{\kappa_{CDW}(0,T)}{T\sigma_{CDW}(0,T)}=\frac{9\zeta(3)}{2\ln(2)}\left(\frac{k_{B}}{e}\right)^{2}\,,
\end{equation}
but with a particular numerical constant $9\zeta(3)/(2\ln(2)) \approx 7.8$.
Note that the order of the limits is essential, otherwise, $\sigma_{CDW}$
is dominated by the interband term in the $|\omega|/(k_{B}T)\gg1$
limit, \[
\lim_{\omega\to0}\sigma_{CDW}(\omega,0)=\frac{v_{F}e^{2}}{2v_{\Delta}}\,,\]
 violating the WF law. We should stress, however that this relation
is typically valid in the DC limit $\omega\to0$, which is well defined
for $\beta|\omega|\ll1$ but not for $\beta|\omega|\gg1$. This is easily
seen by noticing that at $T=0$ the quasiparticle excitation energy
$\omega$ do not have a scale and the AC and DC sectors cannot be
distinguished. For finite $\mu$, it is immediate to check that the WF relation is
verified exactly as in a metal,
\[
\lim_{T\to0}\frac{\kappa_{CDW}(0,T)}{T\sigma_{CDW}(0,T)}=\frac{\pi^{2}}{3}\left(\frac{k_{B}}{e}\right)^{2}\,.\]

\section{Meissner effect}

The non-local electrodynamics is described in the London limit, where
the vector potential function $\mathbf{A}(\mathbf{k})\to\mathbf{A}_{0}=const.$
under the assumption that the field variations are slow in comparison
to the coherence length $\xi$. In this limit, the current $\mathbf{j}$
and the vector potential obey the London equation \[
\langle j_{i}\rangle=j_{\, i}^{CM}+Q_{ij}\, A_{j}\,,\]
valid in the Coulomb gauge $\mathbf{k}\cdot\mathbf{A}=0$, where $\mathbf{j}^{CM}$
is the current due to the momentum of the pair center of mass. For
all purposes, we neglect this effect and consider only the
response to the magnetic field. 

In order to calculate the London kernel $Q_{ij}$, instead of writing the current
density operator (\ref{Ecurrent}), we propose a more general procedure,
extending the CDW band beyond the cone approximation. As in sec. II, 
we start from a CDW Hamiltonian written in terms of an
extended band \begin{equation}
H_{CDW}=\sum_{\mathbf{k},\sigma}\:\Psi_{\sigma}^{\dagger}(\mathbf{k})\left[\epsilon_{\mathbf{k}}\eta_{3}+\Delta_{c\mathbf{k}}\eta_{1}\right]\Psi_{\sigma}(\mathbf{k})\,,\label{NormalCDWBand}\end{equation}
 where $\epsilon_{\mathbf{k}}$ and $\Delta_{c\,\mathbf{k}}$ are
\emph{any} anti-symmetric $k-$functions with respect to a given Fermi
surface node. 

Introducing the magnetic field through the modified Peierls substitution
(\ref{Peirls}), the series expansion of $\epsilon(\mathbf{k}-\frac{e}{c}\tau_{3}\mathbf{A})$
in powers of $\mathbf{A}$ is separated into symmetric and anti-symmetric
terms in $k$, 
\begin{eqnarray*}
\epsilon(\mathbf{k}-\frac{e}{c}\tau_{3}\mathbf{A}) & = &
[\epsilon^{(0)}(\mathbf{k})+\epsilon^{(2)}(\mathbf{k})+\ldots]\tau_{0}+
\\
 &  &
 \qquad[\epsilon^{(1)}(\mathbf{k})+\epsilon^{(3)}(\mathbf{k})+\ldots]\tau_{3}
\\
 & = & [\epsilon_{\mathbf{k}}-\frac{e}{c}\,
 A_{i}\partial^{i}\epsilon_{\mathbf{k}}+\frac{1}{2}\left(\frac{e}{c}\right)^{2}\!\! A_{i}A_{j}\partial^{i}\partial^{j}\epsilon_{\mathbf{k}}]\tau_{0},
\end{eqnarray*}
up to second order in $\mathbf{A}$, where $\partial^{i}\equiv\frac{\partial}{\partial k_{i}}$
defines the momentum derivatives and repeated indexes are to be summed.
The same applies to $\Delta_{c}(\mathbf{k}-\frac{e}{c}\tau_{3}\mathbf{A})$.
Using the abbreviation $\tilde{\mathbf{k}}\equiv\mathbf{k}-\frac{e}{c}\tau_{3}\mathbf{A}$,
the Hamiltonian of the CDW + SC phase with an external magnetic field
is 
\begin{eqnarray*}
H & = &
\sum_{\mathbf{k}}\Psi^{\dagger}(\tilde{\mathbf{k}})\left[\epsilon_{\mathbf{\tilde{\mathbf{k}}}}\tau_{0}\eta_{3}+\Delta_{c\mathbf{\tilde{\mathbf{k}}}}\tau_{0}\eta_{1}\right.
\\
 &  &
 \qquad\qquad\left.+\Delta_{s}\tau_{1}\eta_{2}-\mu\tau_{3}\eta_{0}\right]\Psi(\tilde{\mathbf{k}})\,.
\end{eqnarray*}
The current density operator $\stackrel{\leftrightarrow}{j_{i}}(\mathbf{k)}=-c\nabla_{\mathbf{A}}H$
gives
\begin{eqnarray*}
\stackrel{\leftrightarrow}{j_{i}}(\mathbf{k)} & = &
\Psi^{\dagger}(\tilde{\mathbf{k}})\left[e\left(\partial^{i}\epsilon_{\mathbf{k}}-\frac{e}{c}\,
    A_{j}\partial^{i}\partial^{j}\epsilon_{\mathbf{k}}\right)\tau_{0}\eta_{3}\right.
\\
 & = & \qquad\left.+e\left(\partial^{i}\Delta_{c\mathbf{k}}-\frac{e}{c}\,
     A_{j}\partial^{i}\partial^{j}\Delta_{c\mathbf{k}}\right)\tau_{0}\eta_{1}\right]\Psi_{\tilde{\mathbf{k}}}\,.
\end{eqnarray*}
We calculate the expectation value $\langle\stackrel{\leftrightarrow}{j}\rangle$
up to first order in $\mathbf{A}$ (see details in appendix E), and find that the
London kernel reads, 
\begin{eqnarray}
Q_{ij} & = & \frac{e^{2}}{c}\,\sum_{\mathbf{k}}\sum_{\sigma=\pm1}\nonumber \\
 &  & \left\{
 \frac{\beta}{2}\left[(\partial_{i}\epsilon_{\mathbf{k}})(\partial_{j}\epsilon_{\mathbf{k}})+(\partial_{i}\Delta_{c\mathbf{k}})(\partial_{j}\Delta_{c\mathbf{k}})\right]\right.\nonumber 
\\
 &  & \qquad\qquad\times\textrm{sech}^{2}\left(\frac{\beta
 E_{\mathbf{k},\sigma\mu}}{2}\right)\nonumber 
\\
 &  &
 +\left(\frac{\epsilon_{\mathbf{k}}}{E_{\mathbf{k},\sigma\mu}}\partial_{i}\partial_{j}\epsilon_{\mathbf{k}}+\frac{\Delta_{c\mathbf{k}}}{E_{\mathbf{k},\sigma\mu}}\partial_{i}\partial_{j}\Delta_{c\mathbf{k}}\right)\frac{\sigma\mu+E_{\mathbf{k}}^{*}}{E_{\mathbf{k}}^{*}}\nonumber 
\\
 &  & \qquad\qquad\left.\times\tanh\left(\frac{\beta
 E_{\mathbf{k},\sigma\mu}}{2}\right)\right\} \,,
\label{LK}
\end{eqnarray}
where $E_{\mathbf{k}}^{*}=\sqrt{\epsilon_{\mathbf{k}}^{2}+\Delta_{c\mathbf{k}}^{2}}$
and 
\[
E_{\mathbf{k},\sigma\mu}=\left[\left(\sqrt{\epsilon_{\mathbf{k}}^{2}+\Delta_{c\mathbf{k}}^{2}}+\sigma\mu\right)^{2}+\Delta_{s}^{2}\right]^{\frac{1}{2}}
\]
is the generalized dispersion in the extended CDW band. 

The non-local properties valid in the $q\to0$ limit do not depend on the
details of the cut-off $\Lambda$. For this reason, we are allowed
to take $\Lambda$ to infinity with no further consequences. 
However, the Green functions method leads to some
spurious results in the ultraviolet if we do not take the full Brillouin
zone into account. To see this, consider the illustrative
case of the normal CDW band (\ref{NormalCDWBand}). After a suitable
diagonalization into a particle-hole eigenstate basis with eigenvalues
$\pm E_{\mathbf{k}}^{*}=\pm\sqrt{\epsilon_{\mathbf{k}}^{2}+\Delta_{c\mathbf{k}}^{2}}$
, we may write it into the form: 
\[
H_{CDW}=\sum_{\mathbf{k}}\,
E_{\mathbf{\tilde{\mathbf{k}}}}^{*}\bar{\Psi}^{\dagger}(\tilde{\mathbf{k}})\,\eta_{3}\bar{\Psi}(\tilde{\mathbf{k}})\,.
\]
The London kernel of this problem can be derived directly from Eq. (\ref{LK})
by setting $\Delta_{s}=\mu=0$, ignoring the $\Delta_{c\mathbf{k}}$
terms on it, and performing the substitution $\epsilon_{\mathbf{k}}\to E_{\mathbf{k}}^*$.
It is immediate to see that in this case one has, 
\[
Q_{ij}^{CDW}=A_{j}\sum_{\mathbf{k}}\partial_{i}\left[(\partial_{j}E_{\mathbf{k}}^*)\tanh\left(\frac{\beta
      E_{\mathbf{k}}^*}{2}\right)\right],
\]
resulting in a non-zero surface term for $i=j$, which diverges in the
ultraviolet for any monotonically crescent $E_{\mathbf{k}}^*$. The integrability of the results derived by this method depends
on the introduction of states in the entire Brillouin zone. In particular,
we have that $\langle j_{i}^{CDW}\rangle=0$ (as expected) by
assuming that the surface term cancels in the Brillouin zone because
of its periodicity. In order to fix the spurious divergences, we follow
an argument due to Lifshitz and Pitaevskii \cite{landau}. Considering that the kernel
for $\Delta_{s}=0$ is zero, since no supercurrents are induced by
the magnetic field, there is no physical result in subtracting from the
SC kernel the normal phase kernel, 
\begin{equation}
\langle j_{i}\rangle=\left[Q_{ij}(\Delta_{s})-Q_{ij}(0)\right]A_{j}\,.
\label{Landau}
\end{equation}
We may consider that the kernel above correctly incorporates the Brillouin
zone effects, at least near the phase transition. 

To analyze the spectral weight behavior due to the Meissner effect
within the cone approximation $\epsilon_{\mathbf{k}}\sim v_{F}k_{\perp}$
and $\Delta_{c\mathbf{k}}\sim v_{\Delta}k_{\parallel}$, we calculate
the London equation in two limits, near the normal-SC transition and
at $T=0$. Including the Brillouin zone $[-\frac{\pi}{d},\frac{\pi}{d}]$
in the normal direction to the planes, with $d$ the inter-plane distance,
from Eq. (\ref{LK}) we have 
\[
Q_{\perp}(\Delta_{s})=v_{F}^{2}\frac{\beta}{2}\frac{e^{2}}{c}\,\sum_{\mathbf{k},\sigma=\pm1}\textrm{sech}^{2}\left(\frac{\beta
    E_{\mathbf{k},\sigma\mu}}{2}\right).
\]
At $T=0$, the kernel gives 
\[
Q_{\perp}(\Delta_{s})-Q_{\perp}(0)\stackrel{T\to0}{\longrightarrow}-\frac{|\mu|}{d}\frac{e^{2}v_{F}}{\pi
  v_{\Delta}c}\,,
\]
confirming the anomalous behavior detected by the $f-$sum rule (\ref{sumR})
in the optical conductivity. 

In the opposite limit, for $T\sim T_{c}$, the kernel in the strong
coupling approximation ($\beta_{c}|\mu|\ll1$) gives 
\begin{equation*}
Q_{\perp}(\Delta_{s})-Q_{\perp}(0)\,  \stackrel{T\to T_{c}}{\longrightarrow}
 \,-\frac{\beta_{c}}{4d}\frac{v_{F}e^{2}}{\pi v_{\Delta}c} \left(1+\mu^{2}\frac{\beta_{c}^{2}}{4}\right)\Delta_{s}^{2},
\end{equation*}
in agreement with the mean field result for the penetration depth
$\lambda_{\perp}=\sqrt{c/\{4\pi\left[ Q_{\perp}(0)-Q_\perp(\Delta_s)\right]\}}\propto\Delta_{s}^{-1}$. 

\begin{figure}
\begin{center}$\!\!\!\!\!\!\!\!\!\!\!\!$\includegraphics[%
  scale=0.81]{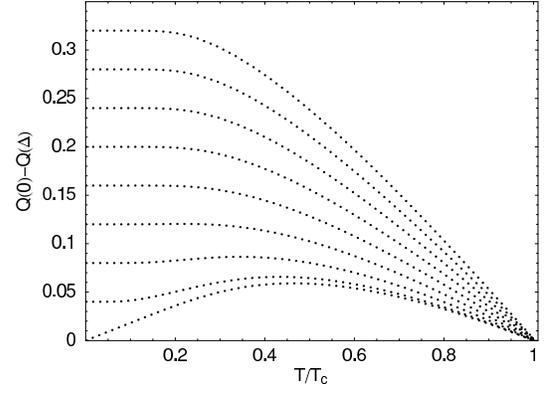}$\qquad\quad$\end{center}

\caption{{\small London kernel dependence with temperature in the cone approximation ($g/g_{c}=1.1$).
Plots for $0\leq|\mu|/\alpha\leq0.16$, from the bottom to the top,
in fixed intervals of $0.02$. $Q(0)-Q(\Delta_{s})$ in units of $e^{2}v_{F}\alpha/(2\pi dv_{\Delta}c)$.}}
\end{figure}

The dependence of the London kernel with $\mu$ and the temperature
is shown in Fig 16. There is a clear suppression of the Meissner effect
in the low temperature region, specially when the density of states
in the Fermi surface nodes is close to zero. As we discussed previously
in sec. VI, the opening of a SC gap in a nodal liquid possibly causes the spectral
shift of high energy states beyond the cone cut-off $\alpha$ in the
CDW band to the infrared. As we mentioned before, the spectral shift
of the states bellow $\alpha$ ($\sim60$ meV) has actually been observed
in the normal CDW phase of the TaSe$_{2}$ \cite{vescoli}.
More experimental studies are required to understand the SC phase
properties in this crystal.

\section{Conclusions}

In this paper we have studied the thermodynamic and transport
properties of a model proposed originally in ref.~[\onlinecite{Neto}]
for the coexistence of a
gapless CDW phase and a $s-$wave superconductor in TMD. 
While the lattice inversion symmetry is broken in the 
CDW distorted phase, as observed
experimentally by neutron diffraction, we propose a pairing ansatz
which also violates the time-reversal symmetry. According to the ansatz,
the pairing of the electrons is mediated by virtual acoustic phonons
\emph{via} a piezoelectric coupling, and the center of mass momentum of
the pairs equals the CDW wavevectors connecting different sheets
of the TMD Fermi surface. 
This additional broken symmetry has dramatic consequences on the spin
exchange interaction and produces 
an anisotropic NMR response along the normal direction to the triple-CDW plane. 
 In contrast to TaSe$_{2}$, the
quasiparticles of NbSe$_{2}$ are well defined in the Fermi-liquid
regime. The SC phase of the NbSe$_2$ has been extensively studied
and indicates that a conventional BCS description is warranted
\cite{craven,garoche,Kobayashi}. 

In contrast to the BCS theory, which is not critical,  
the gap equation (\ref{GapEq}) has a QCP in the critical coupling
$g=g_{c}$ when the system exhibits particle-hole symmetry ($\mu=0$).
When this symmetry is broken, the SC gap $\Delta_{s}$ is strongly
rescaled by $\mu$ as the coupling parameter is modified, and the
QCP is suppressed. The scaling of the quantity $\Delta_{s}/\mu$ follows
two different coupling regimes: (\emph{i}) {}``Fermi liquid'' sector
in weak coupling, for $g<g_{c}$, where $\Delta_{s}/\mu$ flows to
zero as $g\to0$, and (\emph{ii}) strong coupling marginal limit for
$g>g_{c}$, where $\Delta_{s}/|\mu|\gg1$. The specific heat jump is
strongly attenuated in the particle-hole symmetric case 
(where $\Delta C_{V}/C_{n}=0.35$),
because of the low density of states at the Fermi energy. As
expected, in the Fermi liquid regime we recover the jump of
the BCS model $\Delta C_{V}/C_{n}=1.43$.

We have observed several anomalous properties in the transport. Unlike
traditional one-band superconductors, the spectra for optical and
thermal conductivities in the clean limit have an infrared peak due
to the thermal channels of quasi-particle excitation. These channels
involve thermal intraband excitations, promoting the electrons in
the condensate to the empty states of the pocket, 
at the top of the lower band (see Fig. 13). The absorption window for this channel
is limited to the pocket energy $2|\mu|$. A second kind of absorption
channel is due to interband excitations, when a pair is broken as
a result of the absorption of a photon. In this case, the electron
is excited to the upper band, across the optical gap barrier 
$\omega_{0}=2\sqrt{\mu^{2}+\Delta_{s}^{2}}$.
The later type depends on the number of electrons in the condensate
and is more effective at $T=0$, except for a few channels at a given
frequency $\omega_{a}$ which are abruptly suppressed by the temperature
reduction (say, bellow $T_{a}$) because of the optical gap enlargement,
that is, $\omega_{a}<\omega_{0}(T)$ for $T<T_{a}$. The thermal channels
on the contrary vanish at $T=0$ with no exception. 

The optical conductivity has an absorption edge at $\omega_{0}$.
The coherence factors are affected by the broken lattice
inversion symmetry in the CDW phase. The $f-$sum rule reveals an 
anomalous suppression
of the diamagnetic spectral weight, mainly for $\mu=0$. This behavior
is an evidence that there are missing high energy diamagnetic states
in the SC phase, which would be attracted from the bottom to the top
of the lower band by the opening of the SC gap. Close
to the normal-SC phase transition, however, these states can by introduced
by the same procedure that fixes the anomalous divergence of the
London kernel in the ultraviolet, which is due to the absence of the Brillouin
zone periodicity into the calculation.  We have extended the
calculation to a general CDW band where the loss of the crystal 
inversion symmetry is included by assumption. 

In summary, we have presented a complete theory for s-wave superconductivity in nodal
liquids. We have calculated the thermodynamics, the various response functions, and
transport properties of this system and have shown that these quantities
deviate strongly from the same properties in ordinary BCS superconductors 
when there is particle-hole symmetry. We believe our theory can be applied
to some TMD, such as 2H-TaSe$_2$ or 2H-TaS$_2$, and our predictions can
be checked experimentally.

\section*{Acknowledgments}

B. U. and G. G. C. are indebted to E. Miranda for many helpful discussions. 
B. U. acknowledges FAPESP (Funda\c{c}\~ao  de Amparo \`a Pesquisa do Estado
de S\~ao Paulo), project number 00/06881-9, for the financial support.
A.H.C.N. was partially supported through NSF grant 
DMR-0343790.

\appendix

\section{Gap equation}

In this appendix we derive Eq. (\ref{gapSCoupling})$-$(\ref{TcMu}).
Applying the variable substitution $\nu=v_{F}\bar{k}+\sigma\mu$,
the equation (\ref{GapEq}) can be written into the form 
\begin{eqnarray}
1 & = & \frac{gv_{F}}{4\pi
  v_{\Delta}}\sum_{\sigma=\pm1}\int_{0}^{s_{\sigma}}\textrm{d}\bar{k}\,\frac{\bar{k}}{E_{\mathbf{k},\sigma\mu}}\tanh\left(\beta\frac{E_{\mathbf{k},\sigma\mu}}{2}\right)\nonumber 
\\
 & = & \frac{g}{4\pi v_{\Delta}v_{F}}\left\{
  \frac{{4}}{\beta}\,\ln\!\left[\cosh\left(\beta\alpha/2\right)\textrm{sech}\left(\beta\sqrt{\Delta_{s}^{2}+\mu^{2}}/2\right)\right]\right.\nonumber 
\\
 &  &
  \left.+\mu\int_{-\mu}^{\mu}\textrm{d}\nu\,\frac{1}{\sqrt{\nu^{2}+\Delta_{s}^{2}}}\tanh\left(\frac{\beta}{2}\sqrt{\nu^{2}+\Delta_{s}^{2}}\right)\right\}
  .
\label{GapMuT}
\end{eqnarray}

In the $|\mu|/\Delta_{s}\ll1$ limit we find:
\begin{eqnarray*}
1 & = & \frac{g}{4\pi v_{\Delta}v_{F}}\left\{
  \frac{4}{\beta}\,\ln\left[\frac{\cosh\left(\beta\alpha/2\right)}{\cosh\left(\beta\Delta_{s}/2\right)}\right]\right.
\\
 &  &
  \qquad\quad\left.+\frac{\mu^{2}}{\Delta_{s}}\,\tanh\left(\frac{\beta}{2}\Delta_{s}\right)\right\}
  ,
\end{eqnarray*}
that is equivalent to Eq. (\ref{gapSCoupling}). We notice, however,
that the above expression remains valid at $T_{c}$ (\emph{i.e.} for
finite $\mu$ and $\Delta_{s}\to0$) if the strong coupling approximation
$|\mu|/\Delta_{0\mu}\ll1$ is satisfied. 

We define $\alpha=2\pi v_{F}v_{\Delta}/g_{c}$.
Close to $T_{c}$, taking $\Delta_{s}\to0$ we obtain, 
\begin{eqnarray*}
\frac{2g_{c}}{g} & = &
\frac{{4}}{\beta_{c}\alpha}\,\ln\left[\cosh\left(\frac{\beta_{c}\alpha}{2}\right)\right]+\frac{\mu^{2}\beta_{c}}{2\alpha}\,.
\end{eqnarray*}
The critical temperature for $g>g_{c}$ is, 
\begin{equation}
T_{c}=\frac{1}{2k_{B}\ln4}\left[\Delta_0+\sqrt{\Delta_0^{2}+\mu^{2}\ln4}\right],
\label{TcMuAp}
\end{equation}
where $\Delta_0\equiv\Delta_s(T=0,g,\mu=0)=\alpha(1-g_c/g)$.
 The expression that gives the critical dependence of the gap with temperature for $|\mu|/\Delta_{0\mu}\ll 1$  follows directly from the expansion of the gap equation (\ref{GapEq}) in terms of $\beta\Delta_{s}$. 

To calculate the critical temperature in the weak coupling regime,
we take $\Delta_{s}\to0$ in Eq. (\ref{GapMuT}) leading to:
\begin{eqnarray*}
\frac{2g_{c}}{g} & = &
\frac{{4}}{\beta_{c}\alpha}\,\ln\left[\frac{\cosh\left(\beta_{c}\alpha/2\right)}{\cosh\left(\beta_{c}\mu/2\right)}\right]+\frac{2|\mu|}{\alpha}\left[\ln\left(\frac{\beta_{c}|\mu|}{2}\right)\right.
\\
 &  &
 \left.\times\tanh\left(\frac{\beta_{c}\mu}{2}\right)-\int_{0}^{\beta_{c}|\mu|/2}\,\textrm{d}\zeta\,\frac{\ln\zeta}{\cosh^{2}\zeta}\right]
\end{eqnarray*}
after integrating the second term of Eq. (\ref{GapMuT}) by parts.
If $\beta_{c}|\mu| \gtrsim 4$, the integration above can be extended to the interval $[0,\infty]$,
\begin{eqnarray*}
\frac{2g_{c}}{g} & = &
2+\frac{2|\mu|}{\alpha}\left[\ln\left(\frac{\beta_{c}|\mu|\gamma}{\pi}\right)-1\right].
\end{eqnarray*}
In weak coupling ($|\mu|/\Delta_{0\mu}\gg1$) the condition $\beta_{c}|\mu|\gg1$
is easily satisfied. The equation above implies that 
\begin{equation}
T_{c}=\frac{\mu\gamma}{k_{B}\pi}\,\textrm{e}^{\alpha(1-g_{c}/g)|\mu|^{-1}-1}\,,
\label{betaWeak}
\end{equation}
where $\ln(\gamma) \approx 0.577$ is the Euler constant. 

In the weak coupling regime, we can find the gap equation near the
phase transition. For $\beta|\mu|\gg1$, we  use the power series expansion in $\beta\Delta_{s}\ll1$
of the integral \cite{Abrikosov} 
\begin{eqnarray*}
\int_{0}^{\mu}\textrm{d}\nu\,\frac{\tanh\!\left(\frac{\beta}{2}\sqrt{\nu^{2}+\Delta_{s}^{2}}\right)}{\sqrt{\nu^{2}+\Delta_{s}^{2}}}
& \sim &
\int_{0}^{\mu}\textrm{d}\nu\,\frac{1}{\nu}\,\tanh\!\left(\frac{\beta}{2}\nu\right)
\\
 &  & -\frac{7\zeta(3)}{8}\frac{\beta^{2}\Delta_{s}^{2}}{\pi^{2}}
\end{eqnarray*}
Expanding the gap equation (\ref{GapMuT})
in lowest order around $\beta_{c}$, we find 
\begin{equation}
\Delta_{s}(T\to
T_{c},\mu)\stackrel{\beta_{c}\mu\gg1}{\longrightarrow}\frac{1}{\beta_{c}}\left[\frac{7\zeta(3)}{8\pi^{2}}+\frac{1}{2\beta_{c}^{2}\mu^{2}}\right]^{-\frac{1}{2}}\!\!\sqrt{\frac{T_{c}-T}{T_{c}}}\,.
\label{DeltaWeak}
\end{equation}
The weak coupling expansions given above are correct whenever 
$\textrm{tanh}(\beta_c|\mu|/2)\sim1$, or $\beta_c|\mu|\gtrsim4$.

\section{Specific heat}

In this section we calculate explicitly the specific heat jump
in the weak and strong coupling limits.
The entropy of the problem is given by:
\begin{eqnarray*}
S & = &
-k_{B}\,\sum_{\mathbf{k},\gamma,\sigma}\left[(1-n_{\mathbf{k},\sigma\mu}^{\gamma})\,\textrm{ln(}1-n_{\mathbf{k},\sigma\mu}^{\gamma}\textrm{)}\right.
\\
 &  &
 \qquad\qquad\quad\left.+n_{\mathbf{k},\sigma\mu}^{\gamma}\,\textrm{ln}\,
   n_{\mathbf{k},\sigma\mu}^{\gamma}\right],
\end{eqnarray*}
where 
$n_{\mathbf{k},\sigma\mu}^{\gamma}=\left(\textrm{e}^{\gamma\beta
    E_{\mathbf{k},\sigma\mu}} +1\right)^{-1}$ is the Fermi-Dirac
distribution, indexed by $\sigma=\pm1$, and by the two branches of the cone.
$\gamma=\pm1$.  The specific heat yields \cite{tinkham}
\begin{eqnarray}
C_{V} & = & -\beta\frac{\textrm{d}S}{\textrm{d}\beta}\nonumber 
\\
 & = & -k_{B}\beta\sum_{\mathbf{k},\alpha,\sigma}\gamma\,\frac{\partial
   n_{\mathbf{k},\sigma\mu}^{\gamma}}{\partial
   E_{\mathbf{k},\sigma\mu}}\left(E_{\mathbf{k},\sigma\mu}^{2}+\frac{\beta}{2}\frac{\textrm{d}\Delta_{s}^{2}}{\textrm{d}\beta}\right).
\label{CvEq2}
\end{eqnarray}
At the phase transition, the specific heat jump reads:
\begin{eqnarray*}
\Delta C(\beta_{c},\mu) & = &
\lim_{\beta\to\beta_{c}}\left[-k_{B}\frac{\beta_{c}^{2}}{2}\,\frac{\textrm{d}\Delta_{s}^{2}}{\textrm{d}\beta}\,\sum_{\mathbf{k},\gamma,\sigma}\gamma\,\frac{\partial
    n_{\mathbf{k},\sigma\mu}^{\gamma}}{\partial
    E_{\mathbf{k},\sigma\mu}}\right]
\\
 & = & \frac{k_{B}\beta_{c}^{3}}{8\pi
   v_{\Delta}v_{F}}\,\left.\frac{\textrm{d}\Delta_{s}^{2}}{\textrm{d}\beta}\right|_{\beta_{c}}
\\
 &  &
 \times\sum_{\sigma=\pm1}\int_{0}^{\alpha}\textrm{d}\epsilon\,\epsilon\,\textrm{sech}^{2}\!\left(\frac{\beta_{c}(\epsilon+\sigma\mu)}{2}\right).
\end{eqnarray*}
If $\beta_c\alpha\gtrsim 4$, we may extend the integration range to
infinity. This integral can be evaluated in two limits, for $\beta_{c}|\mu|\ll1$
and $\beta_{c}|\mu|\gg1$:
\begin{eqnarray}
\Delta C(\beta,\mu)\!\! & \longrightarrow\!\! & \frac{k_{B}\beta_{c}}{2\pi
  v_{\Delta}v_{F}}\,\left.\frac{\textrm{d}\Delta_{s}^{2}}{\textrm{d}\beta}\right|_{\beta_{c}}\nonumber 
\\
 &  & \times\left\{ \begin{array}{cl}
\ln4+\frac{\beta_{c}^{2}\mu^2}{4} & ,\,\textrm{for}\,\beta_{c}|\mu|\ll1
\\
\\
\beta_{c}|\mu|\, & ,\,\textrm{for}\,\beta_{c}|\mu|\gg1\end{array}\right.
\label{DeltaC}
\end{eqnarray}
From Eq. (\ref{GapTc2}) and (\ref{TcMuAp}), we find:
\begin{eqnarray}
\left.\frac{\textrm{d}\Delta_{s}^{2}}{\textrm{d}\beta}\right|_{\beta_{c}} & = & \left\{ \begin{array}{cl}
\frac{4}{\beta_{c}^{2}}\left(\ln 4+\frac{\beta_{c}\mu^{2}}{4}\right) &
,\,\textrm{for}\,\beta_{c}|\mu|\ll1
\\
\\
\frac{1}{\beta_{c}^{3}}\left(\frac{7\zeta(3)}{8\pi^{2}}+\frac{1}{2\beta_{c}^{2}\mu^{2}}\right)^{-1}
& ,\,\textrm{for}\,\beta_{c}|\mu|\gg1\end{array}\right.\qquad
\label{deltaJump}
\end{eqnarray}

In the normal phase, the specific heat $C_{V\, n}$ is obtained
from Eq. (\ref{CvEq2}),
\begin{eqnarray*}
C_{V\, n}(\beta_{c}) & = & \frac{k_{B}\beta_{c}^{2}}{4\pi
  v_{\Delta}v_{F}}\,\sum_{\sigma}\int_{0}^{\alpha}\textrm{d}\epsilon\,\epsilon\,(\epsilon+\sigma\mu)^{2}
\\
 &  &
  \qquad\times\textrm{sech}^{2}\!\left(\frac{\beta_{c}(\epsilon+\sigma\mu)}{2}\right).
\end{eqnarray*}
Evaluating the integral gives: 
\begin{eqnarray}
C_{V\, n}(\beta_{c}) & \stackrel{}{\longrightarrow} & \frac{k_{B}}{2\pi v_F
  v_\Delta}\frac{1}{\beta_{c}^{2}}\!\times\!\!\left\{ 
\begin{array}{cl}
18\zeta(3) & \,,\,|\mu|\beta_{c}\ll1
\\
\\
\frac{2}{3}\pi^{2}\beta_{c}|\mu| & \,,\,|\mu|
\beta_{c}\gg1\,.
\end{array}\right.
\label{CvMu}
\end{eqnarray}
Combining Eq. (\ref{DeltaC}), (\ref{deltaJump}) and (\ref{CvMu}),
we find
\begin{eqnarray}
\left.\frac{\Delta C_{V}}{C_{n,V}}\right|_{T_{c}} & = & \left\{ 
\begin{array}{cl}
\frac{2\ln4}{9\zeta(3)}\,\left(\ln4+\frac{\beta_{c}^{2}\mu^{2}}{2}\right) &
\,,\,|\mu|\beta_{c}\ll1
\\
\\
\frac{3}{2\pi^{2}}\frac{1}{\frac{7\zeta(3)}{8\pi^{2}}+\frac{1}{2\beta_{c}^{2}\mu^{2}}}
& \,,\,|\mu|\beta_{c}\gg1\,.
\end{array}\right.
\end{eqnarray}

\section{Susceptibilities}

We define the charge and spin susceptibilities from the
imaginary time ordered correlation functions:
\begin{equation}
\chi^{c}(\mathbf{q},i\omega)=-\int_{0}^{\beta}\textrm{d}\tau\,\textrm{e}^{i\omega\tau}\langle
T_{\tau}\left[\rho(\mathbf{q},\tau)\,\rho(-\mathbf{q},0)\right]\rangle
\label{Pi}
\end{equation}
\begin{equation}
\chi_{ab}^{s}(\mathbf{q},i\omega)=-\int_{0}^{\beta}\textrm{d}\tau\,\textrm{e}^{i\omega\tau}\langle
T_{\tau}\left[S_{a}(\mathbf{q},\tau)\,
  S_{b}(-\mathbf{q},0)\right]\rangle,
\label{PiS}
\end{equation}
with $\rho$ and $S_{a}$ respectively as the charge and spin density
operators defined by Eq. (\ref{rhoDens}) and (\ref{S}). 

The optical, thermal and thermoelectric correlation functions are
defined as, 
\begin{equation}
\Pi_{ij}(\mathbf{q},i\omega)=\!-\int_{0}^{\beta}\textrm{d}\tau\,
e^{i\omega\tau}\langle
j_{i}(\mathbf{q},\tau)j_{j}(-\mathbf{q},0)\rangle\label{j-j}
\end{equation}
\begin{equation}
\Pi_{ij}^{EE}(\mathbf{q},i\omega)=\!-\int_{0}^{\beta}\textrm{d}\tau\,
e^{i\omega\tau}\langle
j_{i}^{E}(\mathbf{q},\tau)j_{j}^{E}(-\mathbf{q},0)\rangle
\end{equation}
\begin{equation}
\Pi_{ij}^{E}(\mathbf{q},i\omega)=\!-\int_{0}^{\beta}\textrm{d}\tau\,
e^{i\omega\tau}\langle j_{i}^{E}(\mathbf{q},\tau)j_{j}(-\mathbf{q},0)\rangle,
\label{PiEC}
\end{equation}
where ${\bf j}$ is the electric current operator (\ref{Ecurrent}) and
${\bf j}^{E}$ is the thermal current operator defined by Eq. (\ref{thermalJ}).

\section{Hamiltonian in the Balian-Werthamer space}

In this appendix we discuss Eq. (\ref{HBW}).
The BW space is introduced to extend the pairs space 
(\textbf{$\mathbf{k}\uparrow,-\mathbf{k}\downarrow$})
to a larger one where the spin and momentum degrees of freedom are
decoupled. The procedure rests on {}``duplicating'' the Hamiltonian (keeping
it invariant by summing in half Brillouin zone), interchange
the order of the $\psi$ fermionic operators in the duplicated term
and explore the symmetry under the $\mathbf{k}\to-\mathbf{k}$ exchange
in the $k-$sum. The CDW Hamiltonian in the BW space reads
\begin{eqnarray}
H_{CDW} & = & \!\!\!\sum_{\mathbf{k},\sigma,a,b}\,
v_{F}\,\psi_{a\,\sigma}^{\dagger}(\mathbf{k})\,\bar{\mathbf{k}}\cdot\vec{\eta}^{a\,
  b}\psi_{b\,\sigma}(\mathbf{k})\nonumber 
\\
 & = &
 \frac{v_{F}}{2}\,\sum_{\mathbf{k},a,b}\bar{\mathbf{k}}\cdot\left[\psi_{a\,\uparrow}^{\dagger}(\mathbf{k})\,\vec{\eta}^{a\, b}\psi_{b\,\uparrow}(\mathbf{k})\right.\nonumber 
\\
 &  & +\psi_{a\,\downarrow}^{\dagger}(\mathbf{k})\,\vec{\eta}^{a\,
   b}\psi_{b\,\downarrow}(\mathbf{k})+\psi_{b\,\uparrow}(-\mathbf{k})\,\vec{\eta}^{b\, a}\psi_{a\,\uparrow}^{\dagger}(-\mathbf{k})\nonumber 
\\
 &  & \qquad\qquad\left.+\psi_{b\,\downarrow}(-\mathbf{k})\,\vec{\eta}^{b\,
     a}\psi_{a\,\downarrow}^{\dagger}(-\mathbf{k})\right]\nonumber 
\\
 & = &
 \sum_{\mathbf{k}\in\frac{1}{2}\textrm{B.Z.}}v_{F}\Psi^{\dagger}(\mathbf{k})\,\sigma_{0}\tau_{0}\vec{\eta}\cdot\bar{\mathbf{k}}\,\Psi(\mathbf{k})\,,
\label{HCDWBW}
\end{eqnarray}
by the definition of the BW spinor (\ref{BWS}). 

The chemical potential term (\ref{Hmu}) can also be written as 
$-\mu\sum_{\mathbf{k}\in\frac{1}{2}\textrm{B.Z.}}\Psi^{\dagger}(\mathbf{k})\,\sigma_{0}\tau_{3}\eta_{0}\,\Psi(\mathbf{k})$.
The pairing term can also be obtained with the use of the antisymmetric
property of the Pauli matrix $\eta_{2}$ under the transposition
$\eta_{2}^{a\, b}\to-\eta_{2}^{b\, a}$ , namely,
\begin{eqnarray}
H_{P} & = &
\sum_{\mathbf{k},a,b}\,\Delta_{s}\,\psi_{a\,\uparrow}^{\dagger}(\mathbf{k})\,\eta_{2}^{a\,
  b}\psi_{b\,\downarrow}^{\dagger}(-\mathbf{k})+h.c.\nonumber 
\\
 & = &
 \frac{1}{2}\,\sum_{\mathbf{k},a,b}\,\Delta_{s}\left[\psi_{a\,\uparrow}^{\dagger}(\mathbf{k})\,\eta_{2}^{a\, b}\psi_{b\,\downarrow}^{\dagger}(-\mathbf{k})\right.\qquad\nonumber 
\\
 &  & \left.+\psi_{a\downarrow}(-\mathbf{k})\,\eta_{2}^{a\,
     b}\psi_{b\uparrow}(\mathbf{k})+\psi_{b\,\downarrow}^{\dagger}(\mathbf{k})\,\eta_{2}^{b\, a}\psi_{a\,\uparrow}^{\dagger}(-\mathbf{k})\right]\nonumber 
\\
 &  & \qquad\qquad\left.+\psi_{b\,\uparrow}(-\mathbf{k})\,\eta_{2}^{b\,
     a}\psi_{a\,\downarrow}(\mathbf{k})\right]\nonumber 
\\
 & = &
 -\sum_{\mathbf{k}\in\frac{1}{2}\textrm{B.Z.}}\Delta_{s\,}\Psi^{\dagger}(\mathbf{k})\,\sigma_{3}\tau_{1}\eta_{2}\,\Psi(\mathbf{k})\,.
\label{Int}
\end{eqnarray}
 
\section{London kernel}

In this appendix, we evaluate the London kernel (\ref{LK}).
It can be derived from the calculation of the expectation value of the current
density operator, 
\begin{eqnarray}
\langle \stackrel{\leftrightarrow}{j}_{i}\rangle(\mathbf{k)} & \!\!= &
\!\!\textrm{Tr}\sum_{\mathbf{k}}\left[e\left(\partial_{i}\epsilon_{\mathbf{k}}-\frac{e}{c}A^{j}\partial_{i}\partial_{j}\epsilon_{\mathbf{k}}\right)\right.\nonumber 
\\
 &  &
 \qquad\qquad\qquad\times\langle\Psi^{\dagger}(\tilde{\mathbf{k}})\,\tau_{0}\eta_{3}\Psi(\tilde{\mathbf{k}})\rangle\nonumber 
\\
 &  &
 \left.+e\left(\partial_{i}\Delta_{c\mathbf{k}}-\frac{e}{c}A^{j}\partial_{i}\partial_{j}\Delta_{c\mathbf{k}}\right)\langle\Psi^{\dagger}(\tilde{\mathbf{k}})\,\tau_{0}\eta_{1}\Psi(\tilde{\mathbf{k}})\rangle\right]\nonumber 
\\
\label{<JMu>}
\end{eqnarray}
in first order in $\mathbf{A}$, where in our definition $\tilde{\mathbf{k}}=\mathbf{k}-\frac{e}{c}\tau_{3}\mathbf{A}$.
Expanding the Green function $\stackrel{\leftrightarrow}{G}(i\omega_{n},\tilde{\mathbf{k}})=(i\omega_{n}-\stackrel{\leftrightarrow}{\omega}_{\tilde{\mathbf{k}}})^{-1}$
up to leading order, 
\begin{eqnarray*}
\textrm{Tr}\langle\Psi^{\dagger}(\tilde{\mathbf{k}})\,\tau_{\mu}\eta_{\nu}\Psi(\tilde{\mathbf{k}})\rangle
& = &
\frac{1}{\beta}\textrm{Tr}\sum_{\omega_{n}}\tau_{\mu}\eta_{\nu}\stackrel{{\leftrightarrow}}{G}_{0}\left[1-\frac{e}{c}\stackrel{{\leftrightarrow}}{G}_{0}\right.
\\
 &  &
 \quad\left.\times\left(\partial_{i}\epsilon_{\mathbf{k}}\tau_{0}\eta_{3}+\partial_{i}\Delta_{c\mathbf{k}}\tau_{0}\eta_{1}\right)A_{i}\right]\,,
\end{eqnarray*}
where $\stackrel{\leftrightarrow}{G}_{0}$ is the Green function
(\ref{GreenF1}).
The zeroth order terms are: 
\begin{eqnarray}
\textrm{Tr}\langle\Psi^{\dagger}(\tilde{\mathbf{k}})\,\tau_{0}\eta_{3}\Psi(\tilde{\mathbf{k}})\rangle_{0}
& = &
\frac{1}{\beta}\textrm{Tr}\sum_{\omega_{n}}\tau_{0}\eta_{3}\stackrel{\leftrightarrow}{G}_{0}
\nonumber 
\\
 & = &
 \epsilon_{\mathbf{k}}\sum_{\sigma=\pm1}\frac{\mu\sigma +E_{\mathbf{k}}^{*}}{E_{\mathbf{k}}^{*}\, E_{\mathbf{k},\sigma\mu}}
\nonumber 
\\
 &  &
 \times\left[n(E_{\mathbf{k},\sigma\mu})-n(-\mathbf{\textrm{$E$}_{\mathbf{k},\sigma\mu}})\right],\qquad
\label{TrMu1}
\end{eqnarray}
and
\begin{eqnarray}
\textrm{Tr}\langle\Psi^{\dagger}(\tilde{\mathbf{k}})\,\tau_{0}\eta_{1}\Psi(\tilde{\mathbf{k}})\rangle_{0}
& = &
\Delta_{c\mathbf{k}}\sum_{\sigma=\pm1}\frac{\mu\sigma+E_{\mathbf{k}}^{*}}{E_{\mathbf{k}}^{*}E_{\mathbf{k},\sigma\mu}}\nonumber 
\\
 &  &
 \times\left[n(E_{\mathbf{k},\sigma\mu})-n(-\mathbf{\textrm{$E$}_{\mathbf{k},\sigma\mu}})\right],\qquad
\label{TrMu2}
\end{eqnarray}
where $E_{\mathbf{k}}^{*}\equiv\sqrt{\epsilon_{\mathbf{k}}^{2}+\Delta_{c\mathbf{k}}^{2}}$
and \[
E_{\mathbf{k},\sigma\mu}=[(E_{\mathbf{k}}^{*}+\sigma\mu)^{2}+\Delta_{s}^{2}]^{\frac{1}{2}}.\]
 
To first order we find, after a straightforward algebra: 
\begin{eqnarray}
\textrm{Tr}\langle\Psi^{\dagger}(\tilde{\mathbf{k}})\,\tau_{0}\eta_{3}\Psi(\tilde{\mathbf{k}})\rangle_{1}
& = &
-\frac{e}{c}A_{i}\partial_{i}\epsilon_{\mathbf{k}}\,\frac{1}{\beta}\sum_{\omega_{n}}\textrm{Tr}\left[(\stackrel{\leftrightarrow}{G}_{0})^{2}\right]\nonumber 
\\
 & = &
 -\frac{e}{c}\sum_{\sigma=\pm1}A^{i}(\partial_{i}\epsilon_{\mathbf{k}})\nonumber 
\\
 &  & \times\frac{\partial}{\partial
   E_{\sigma\mu}}\left[n(E_{\mathbf{k},\sigma\mu})-n(-E_{\mathbf{k},\sigma\mu})\right].\nonumber 
\\
\label{TrMu3}
\end{eqnarray}
and 
\begin{eqnarray}
\textrm{Tr}\langle\Psi^{\dagger}(\tilde{\mathbf{k}})\,\tau_{0}\eta_{1}\Psi(\tilde{\mathbf{k}})\rangle_{1}
& = &
-\frac{e}{c}\sum_{\sigma=\pm1}A^{i}(\partial_{i}\Delta_{c\mathbf{k}})\nonumber 
\\
 &  & \times\frac{\partial}{\partial
 E_{\sigma\mu}}\left[n(E_{\mathbf{k},\sigma\mu})-n(-E_{\mathbf{k},\sigma\mu})\right].\nonumber 
\\
\label{TrMu4}
\end{eqnarray}

The London kernel (\ref{LK}) follows from the direct substitution
of Eq. (\ref{TrMu1}), (\ref{TrMu2}), (\ref{TrMu3}) and (\ref{TrMu4})
into Eq. (\ref{<JMu>}), and by noting that the zero order current term,
\begin{eqnarray*}
\langle j\rangle_{0} & = & \sum_{\sigma=\pm1}\sum_{\mathbf{k}}\,
e\left(\epsilon_{\mathbf{k}}\partial_{i}\epsilon_{\mathbf{k}}+\Delta_{c\mathbf{k}}\partial_{i}\Delta_{c\mathbf{k}}\right)\frac{\mu\sigma+E_{\mathbf{k}}^{*}}{E_{\mathbf{k}}^{*}\,
  E_{\mathbf{k},\sigma\mu}}
\\
 &  &
 \quad\times\left[n(E_{\mathbf{k},\sigma\mu})-n(-\mathbf{\textrm{$E$}_{\mathbf{k},\sigma\mu}})\right]\end{eqnarray*}
vanishes by symmetry when integrated over ${\bf k}$.

\end{document}